\documentclass[a4paper,11pt]{article}

\usepackage[top=1in,right=0.8in,left=0.8in,bottom=1.0in]{geometry}
 
\usepackage{graphics,graphicx}
\usepackage{subfig}
\usepackage{float}
\usepackage{booktabs}
\usepackage{longtable}
\usepackage{supertabular}
\usepackage{tabularx}
\usepackage{multirow}
\usepackage{array}
\usepackage{tabu}
\usepackage{mathtools}
\usepackage{hyperref}
\usepackage{mathrsfs}

\usepackage{amsmath}
\usepackage{amsfonts}
\usepackage{amssymb}
\usepackage{siunitx} 

\usepackage{enumerate}
\usepackage{sectsty}
\usepackage[affil-it,auth-sc]{authblk}
\usepackage{color}
\usepackage[dvipsnames]{xcolor}

\sectionfont{\normalsize}
\subsectionfont{\normalsize}

\usepackage{tikz}
\usetikzlibrary{arrows,decorations,backgrounds,shapes, snakes}
\usetikzlibrary{positioning}
\usetikzlibrary{matrix} 
\usepackage{varwidth}
\usepackage[most]{tcolorbox}
\usetikzlibrary{shapes.geometric,arrows.meta,decorations.markings}
\usepackage{fancyhdr}            
\fancypagestyle{plain}{%
  \fancyfoot[CE,CO]{\thepage}       
  \fancyhead[CE,CO]{\textcolor[rgb]{0.64,0.15,0.15}{Published in \emph{European Journal of Mechanics - A/Solids}  \textbf{105}: 105273 (2024) 
  \\
doi: https://doi.org/10.1016/j.euromechsol.2024.105273}}
\setlength{\headheight}{14.5pt}}

\begin{document}

\newcommand{\singlespace}{\baselineskip=12pt\lineskiplimit=0pt\lineskip=0pt}
\def\ds{\displaystyle}

\tikzstyle{every picture}+=[remember picture]

\newcommand{\beq}{\begin{equation}}
\newcommand{\eeq}{\end{equation}}
\newcommand{\lb}{\label}
\newcommand{\ph}{\phantom}
\newcommand{\beqar}{\begin{eqnarray}}
\newcommand{\eeqar}{\end{eqnarray}}
\newcommand{\barr}{\begin{array}}
\newcommand{\earr}{\end{array}}
\newcommand{\jump}{\parallel}
\newcommand{\Ehat}{\hat{E}}
\newcommand{\That}{\hat{\bf T}}
\newcommand{\Ahat}{\hat{A}}
\newcommand{\chat}{\hat{c}}
\newcommand{\shat}{\hat{s}}
\newcommand{\khat}{\hat{k}}
\newcommand{\muhat}{\hat{\mu}}
\newcommand{\mc}{M^{\scriptscriptstyle C}}
\newcommand{\mei}{M^{\scriptscriptstyle M,EI}}
\newcommand{\mec}{M^{\scriptscriptstyle M,EC}}
\newcommand{\hbeta}{{\hat{\beta}}}
\newcommand{\rec}[2]{\left( #1 #2 \ds{\frac{1}{#1}}\right)}
\newcommand{\rep}[2]{\left( {#1}^2 #2 \ds{\frac{1}{{#1}^2}}\right)}
\newcommand{\derp}[2]{\ds{\frac {\partial #1}{\partial #2}}}
\newcommand{\derpn}[3]{\ds{\frac {\partial^{#3}#1}{\partial #2^{#3}}}}
\newcommand{\dert}[2]{\ds{\frac {d #1}{d #2}}}
\newcommand{\dertn}[3]{\ds{\frac {d^{#3} #1}{d #2^{#3}}}}
\newcommand{\ct}{\captionof{table}}
\newcommand{\cf}{\captionof{figure}}

\def\c{{\circ}}
\def\bob{{\, \underline{\overline{\otimes}} \,}}
\def\ob{{\, \underline{\otimes} \,}}
\def\scalp{\mbox{\boldmath$\, \cdot \, $}}
\def\gdp{\makebox{\raisebox{-.215ex}{$\Box$}\hspace{-.778em}$\times$}}
\def\daa{\makebox{\raisebox{-.050ex}{$-$}\hspace{-.550em}$: ~$}}
\def\mK{\mbox{${\mathcal{K}}$}}
\def\cK{\mbox{${\mathbb {K}}$}}

\def\Xint#1{\mathchoice
   {\XXint\displaystyle\textstyle{#1}}%
   {\XXint\textstyle\scriptstyle{#1}}%
   {\XXint\scriptstyle\scriptscriptstyle{#1}}%
   {\XXint\scriptscriptstyle\scriptscriptstyle{#1}}%
   \!\int}
\def\XXint#1#2#3{{\setbox0=\hbox{$#1{#2#3}{\int}$}
     \vcenter{\hbox{$#2#3$}}\kern-.5\wd0}}
\def\ddashint{\Xint=}
\def\fpint{\Xint=}
\def\dashint{\Xint-}
\def\cpvint{\Xint-}
\def\intl{\int\limits}
\def\cpvintl{\cpvint\limits}
\def\fpintl{\fpint\limits}
\def\ointl{\oint\limits}
\def\bA{{\bf A}}
\def\ba{{\bf a}}
\def\bB{{\bf B}}
\def\bb{{\bf b}}
\def\bc{{\bf c}}
\def\bC{{\bf C}}
\def\bD{{\bf D}}
\def\bE{{\bf E}}
\def\be{{\bf e}}
\def\bbf{{\bf f}}
\def\bF{{\bf F}}
\def\bG{{\bf G}}
\def\bg{{\bf g}}
\def\bi{{\bf i}}
\def\bH{{\bf H}}
\def\bK{{\bf K}}
\def\bL{{\bf L}}
\def\bM{{\bf M}}
\def\bN{{\bf N}}
\def\bn{{\bf n}}
\def\bm{{\bf m}}
\def\b0{{\bf 0}}
\def\bo{{\bf o}}
\def\bX{{\bf X}}
\def\bx{{\bf x}}
\def\bP{{\bf P}}
\def\bp{{\bf p}}
\def\bQ{{\bf Q}}
\def\bq{{\bf q}}
\def\bR{{\bf R}}
\def\bS{{\bf S}}
\def\bs{{\bf s}}
\def\bT{{\bf T}}
\def\bt{{\bf t}}
\def\bU{{\bf U}}
\def\bu{{\bf u}}
\def\bv{{\bf v}}
\def\bw{{\bf w}}
\def\bW{{\bf W}}
\def\by{{\bf y}}
\def\bz{{\bf z}}
\def\T{{\bf T}}
\def\Te{\textrm{T}}
\def\Id{{\bf I}}
\def\bxi{\mbox{\boldmath${\xi}$}}
\def\balpha{\mbox{\boldmath${\alpha}$}}
\def\bbeta{\mbox{\boldmath${\beta}$}}
\def\bepsilon{\mbox{\boldmath${\epsilon}$}}
\def\bvarepsilon{\mbox{\boldmath${\varepsilon}$}}
\def\bomega{\mbox{\boldmath${\omega}$}}
\def\bphi{\mbox{\boldmath${\phi}$}}
\def\bsigma{\mbox{\boldmath${\sigma}$}}
\def\bfeta{\mbox{\boldmath${\eta}$}}
\def\bDelta{\mbox{\boldmath${\Delta}$}}
\def\btau{\mbox{\boldmath $\tau$}}
\def\tr{{\rm tr}}
\def\dev{{\rm dev}}
\def\div{{\rm div}}
\def\Div{{\rm Div}}
\def\Grad{{\rm Grad}}
\def\grad{{\rm grad}}
\def\Lin{{\rm Lin}}
\def\Sym{{\rm Sym}}
\def\Skw{{\rm Skew}}
\def\abs{{\rm abs}}
\def\Re{{\rm Re}}
\def\Im{{\rm Im}}
\def\capB{\mbox{\boldmath${\mathsf B}$}}
\def\capC{\mbox{\boldmath${\mathsf C}$}}
\def\capD{\mbox{\boldmath${\mathsf D}$}}
\def\capE{\mbox{\boldmath${\mathsf E}$}}
\def\capG{\mbox{\boldmath${\mathsf G}$}}
\def\tcapG{\tilde{\capG}}
\def\capH{\mbox{\boldmath${\mathsf H}$}}
\def\capK{\mbox{\boldmath${\mathsf K}$}}
\def\capL{\mbox{\boldmath${\mathsf L}$}}
\def\capM{\mbox{\boldmath${\mathsf M}$}}
\def\capR{\mbox{\boldmath${\mathsf R}$}}
\def\capW{\mbox{\boldmath${\mathsf W}$}}

\def\i{\mbox{${\mathrm i}$}}
\def\mC{\mbox{\boldmath${\mathcal C}$}}
\def\mB{\mbox{${\mathcal B}$}}
\def\mE{\mbox{${\mathcal{E}}$}}
\def\mL{\mbox{${\mathcal{L}}$}}
\def\mK{\mbox{${\mathcal{K}}$}}
\def\mV{\mbox{${\mathcal{V}}$}}
\def\C{\mbox{\boldmath${\mathcal C}$}}
\def\E{\mbox{\boldmath${\mathcal E}$}}

\def\AAM{{\it Advances in Applied Mechanics }}
\def\ACME{{\it Arch. Comput. Meth. Engng.}}
\def\ARMA{{\it Arch. Rat. Mech. Analysis}}
\def\AMR{{\it Appl. Mech. Rev.}}
\def\ASCEEM{{\it ASCE J. Eng. Mech.}}
\def\ACTA{{\it Acta Mater.}}
\def\CMAME {{\it Comput. Meth. Appl. Mech. Engrg.}}
\def\CRAS{{\it C. R. Acad. Sci. Paris}}
\def\CRM{{\it Comptes Rendus M\'ecanique}}
\def\EFM{{\it Eng. Fracture Mechanics}}
\def\EJMA{{\it Eur.~J.~Mechanics-A/Solids}}
\def\IJES{{\it Int. J. Eng. Sci.}}
\def\IJF{{\it Int. J. Fracture}}
\def\IJMS{{\it Int. J. Mech. Sci.}}
\def\IJNAMG{{\it Int. J. Numer. Anal. Meth. Geomech.}}
\def\IJP{{\it Int. J. Plasticity}}
\def\IJSS{{\it Int. J. Solids Structures}}
\def\IngA{{\it Ing. Archiv}}
\def\JAM{{\it J. Appl. Mech.}}
\def\JAP{{\it J. Appl. Phys.}}
\def\JAE{{\it J. Aerospace Eng.}}
\def\JE{{\it J. Elasticity}}
\def\JM{{\it J. de M\'ecanique}}
\def\JMPS{{\it J. Mech. Phys. Solids}}
\def\JSV{{\it J. Sound and Vibration}}
\def\MACRO{{\it Macromolecules}}
\def\MMT{{\it Mech. Mach. Th.}}
\def\MOM{{\it Mech. Materials}}
\def\MMS{{\it Math. Mech. Solids}}
\def\MMT{{\it Metall. Mater. Trans. A}}
\def\MPCPS{{\it Math. Proc. Camb. Phil. Soc.}}
\def\MSE{{\it Mater. Sci. Eng.}}
\def\NATURE{{\it Nature}}
\def\NATUREM{{\it Nature Mater.}}
\def\PHIL{{\it Phil. Trans. R. Soc.}}
\def\PMPS{{\it Proc. Math. Phys. Soc.}}
\def\PNAS{{\it Proc. Nat. Acad. Sci.}}
\def\PRE{{\it Phys. Rev. E}}
\def\PRL{{\it Phys. Rev. Letters}}
\def\PRSL{{\it Proc. R. Soc.}}
\def\RIIT{{\it Rozprawy Inzynierskie - Engineering Transactions}}
\def\ROCK{{\it Rock Mech. and Rock Eng.}}
\def\QAM{{\it Quart. Appl. Math.}}
\def\QJMAM{{\it Quart. J. Mech. Appl. Math.}}
\def\SCIENCE{{\it Science}}
\def\SCRMAT{{\it Scripta Mater.}}
\def\SM{{\it Scripta Metall.}}
\def\ZAMM{{\it Z. Angew. Math. Mech.}}
\def\ZAMP{{\it Z. Angew. Math. Phys.}}
\def\ZVDI{{\it Z. Verein. Deut. Ing.}}

\def\salto#1#2{
[\mbox{\hspace{-#1em}}[#2]\mbox{\hspace{-#1em}}]}

\renewcommand\Affilfont{\itshape}
\setlength{\affilsep}{1em}
\renewcommand\Authsep{, }
\renewcommand\Authand{ and }
\renewcommand\Authands{ and }
\setcounter{Maxaffil}{2}

\title{The elastica sling}
\author[]{A. Cazzolli}
\author[]{F. Dal Corso\footnote{Corresponding author: francesco.dalcorso@unitn.it}}
\affil[]{DICAM, University of Trento, via~Mesiano~77, I-38123 Trento, Italy}
%
%
%
%
%
%
\date{}
\maketitle

\begin{abstract}
The nonlinear mechanics of a flexible elastic  rod constrained at its edges by a pair of sliding sleeves is analyzed. The planar equilibrium configurations of this variable-length  elastica are found to have shape defined only by the  inclination of the two constraints, while their distance  is responsible only for scaling the size. By extending the theoretical stability criterion available for systems under isoperimetric constraints to the case of variable domains, the existence of no more than one stable equilibrium solution is revealed. The set of sliding sleeves' inclination pairs for which the stability is lost are identified. Such critical conditions allow the  indefinite ejection of the flexible rod from the sliding sleeves, thus realizing an elastica sling. Finally, the theoretical findings are  validated by experiments on a physical prototype. The present results lead to a novel actuation principle that may find application as a mechanism in 
energy harvesting, wave mitigation devices, and soft robotic locomotion.

\end{abstract}


\section{Introduction}

Examples of highly efficient flexible devices can be found throughout history: the ancient war engines made of wood and ropes for throwing stones and arrows (namely, the catapult and the ballista). The performance of the Greek catapult was largely lost a few centuries after the Romans copied the weapon \cite{payne}. Furthermore, the design of a revised version of the catapult, in which the throwing arm is subject to large rotations, is the so-called ‘da Vinci catapult’ (drawn in “Il Codice Atlantico”). Other examples of fascinating efficient compliant mechanisms can be found in nature. Many animals (e.g. snakes and fishes) continuously and deeply change their shape to provide locomotive forces on solid surfaces and in fluids.

The field of structural mechanics is experiencing a second lease on life in the last decade, as the use of extremely deformable structures has become the new paradigm in the design of reconfigurable, wave mitigation, energy harvesting, and actuation mechanisms. Indeed, while in the past the nonlinear regime of structures was avoided as a safety measure in the classical mechanical design,  today  it drives the main strategy for realizing devices with novel and unexpected mechanical responses \cite{kochmann, reis}.

Soft robotics is a relatively young  field that has emerged with the aim of overcoming the inherent limitations of traditional robots by replacing stiff components with highly compliant ones. In addition to dramatically reducing  safety issues in the interaction with humans and objects, compliance is also exploited to provide different locomotion modalities \cite{calisti}. Recently, actuation through an impulsive motion generated by snapping instabilities has been considered for realizing wearable devices for human joint impedance estimation \cite{yagi}, tasks of inflation/deflation routine  of a fluid-filled cavity \cite{arakawa}, and  high performance jumping  \cite{misu0,misu} and swimming \cite{yamada} robots.

More in general, the design of mechanical metamaterials, materials displaying features not achievable from the  classical ones, has been recently enhanced by considering the influence of  structural  flexibility \cite{bertoldirev} towards  multistability \cite{abbassi, amor, maurini, sano},  reconfigurable structures \cite{filipov, liu}, wave guiding  \cite{bordiga,carta, garau} and unidirectional wave propagation \cite{nadkarni, raney}.

Within this renewed interest in nonlinear structural mechanics, the attention of scientists and researchers has also been  drawn by  structures with variable domain. The most famous example is the  concentric tube robot, commonly used for minimally invasive surgeries \cite{alfalahi, mahoney, renda}. More specifically, a variable domain structure is realized whenever one element can slide along another providing a variation in its overall \lq exterior' length or surface area. 

The framework of configurational mechanics, introduced by Eshelby \cite{eshelby} to model defect motion within solids, has  recently been extended to structures subject to sliding sleeve constraints \cite{bigoniconf}. The existence of an outward reaction acting along the sliding direction was demonstrated and  shown  to be quadratic in the curvature value of the  flexible element  at the exit of the sleeve.
Since then, this concept has been exploited
in bifurcation problems \cite{bigoniblade,bosiinj,bosirestab,liakou}, nonlinear dynamics \cite{armaniniconf,koutso}, wave propagation \cite{dalcorsonested}, and to design novel actuation mechanisms, such as the torsional  \cite{bigonitorsional} and the frictionless curved channel  \cite{dalcorsochannel} locomotions. The introduced theoretical framework
led to a new interpretation of dislocation motion \cite{ballarini},
blistering \cite{goldberg, wang}, delamination \cite{venkatadri}, and
penetration \cite{wen} processes. In addition, the first numerical models for the treatment of variable domain structures have recently appeared
\cite{Han2022, Han2023a, Han2023b}.

\begin{figure}[!h]
\renewcommand{\figurename}{\footnotesize{Fig.}}
    \begin{center}
   \includegraphics[width=.9\textwidth]{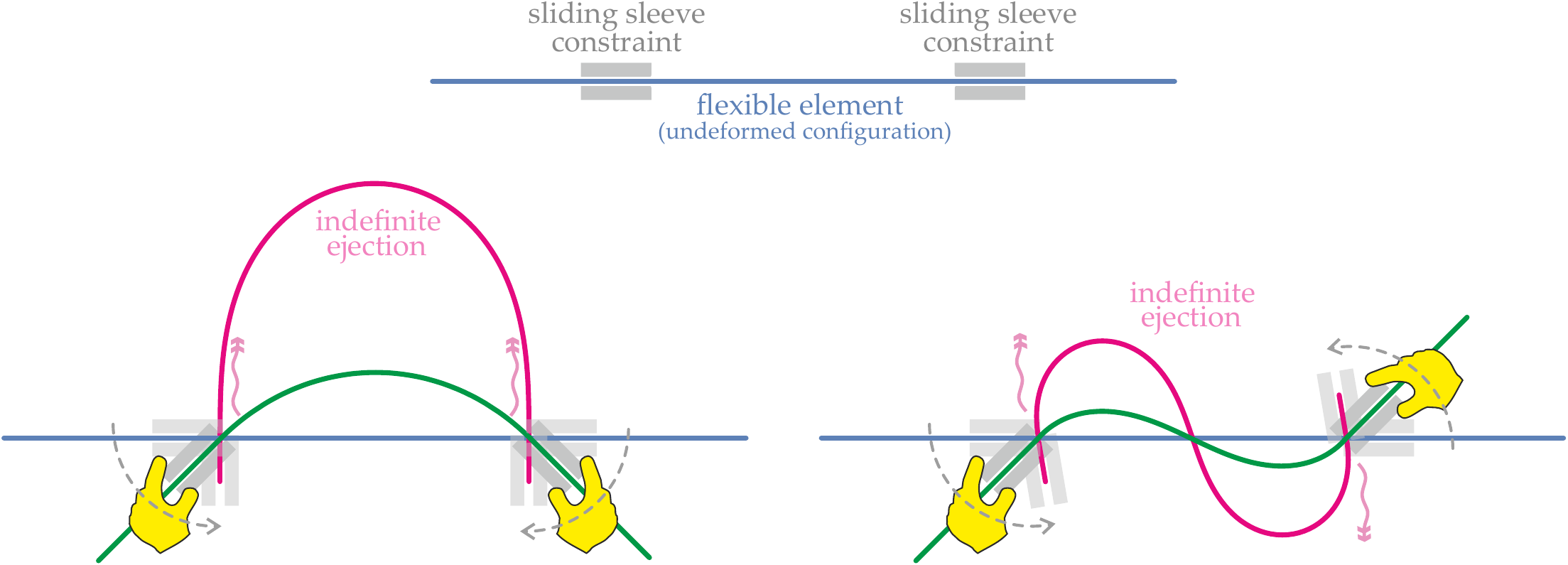}
    \caption{\footnotesize Sketch of the \lq \emph{elastica sling}' realized through the rotation of two sliding sleeves constraining a flexible elements. The system  in its undeformed configuration (above) and  during two different rotations paths (bottom).  More specifically, a symmetric and an antisymmetric evolution is  reported on the bottom left and on the bottom right, respectively. With reference to the nomenclature provided in the following, the symmetric evolution is referred to $\theta_1=-\theta_2=\overline{\theta}$ (namely, $\theta_A=0$), while the antisymmetric one to $\theta_1=\theta_2=\overline{\theta}$ (namely, $\theta_S=0$). Two deformed configurations are reported, associated with $\overline{\theta}=\pi/4$ and with $\overline{\theta}=\overline{\theta}_{cr}$, the critical value providing  the indefinite ejection of the flexible element from the sliding sleeve constraints, which is $\overline{\theta}_{cr}=\pi/2$ (left) and $\overline{\theta}_{cr}\approx 1.106\pi/2$ (right).}
    \label{fig0}
    \end{center}
\end{figure}

The nonlinear mechanics of a flexible elastic  rod constrained at its edges by a pair of sliding sleeves is analyzed in this article (Fig. \ref{fig0}, above). 
In Sect. 2, the mechanical model is formulated and the equilibrium  together with the second and third variations are derived through a variational approach. The planar equilibrium configurations of this variable-length  elastica are obtained in Sect. 3 and found to have shape defined only by the  inclination of the two constraints, while their distance  is responsible only for scaling of size. A theoretical stability criterion is established in Sect. 4 by  extending to the case of variable domains a previous method available for fixed-length systems under isoperimetric constraints.
The developed theoretical framework is exploited in Sect. 5 to show that  there is no more than one stable equilibrium solution for each  pair of inclinations and to define the corresponding values  at which  stability is lost. The  indefinite ejection of the flexible rod from the sliding sleeves occurs at these critical conditions, thus realizing an \lq \emph{elastica sling}' (Fig. \ref{fig0}, bottom). These theoretical findings are finally validated (at the \emph{Instabilities Lab} of
the University of Trento) through experiments on a physical prototype.

The present analysis provides a nonlinear  mechanism that can be used as a building block in the design of novel strategies for 
energy harvesting, wave mitigation devices, and soft robotic locomotion.

\section{Formulation}

\subsection{Geometry and kinematics}

\begin{figure}[!h]
\renewcommand{\figurename}{\footnotesize{Fig.}}
    \begin{center}
   \includegraphics[width=1\textwidth]{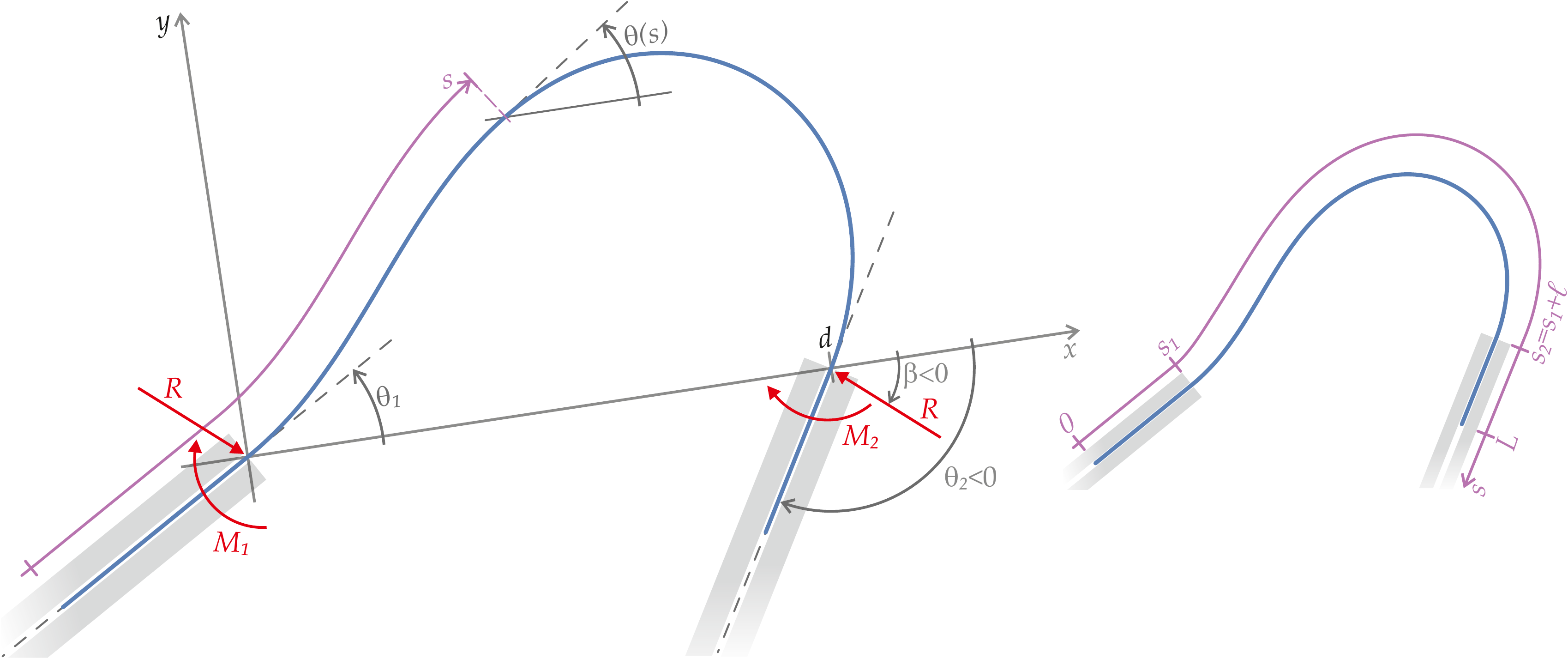}
    \caption{\footnotesize The \lq \emph{elastica sling}' realized through a flexible rod of uniform bending stiffness $B$ and length $L$, constrained at each edge by a different sliding sleeve. (Left)  The configuration of the system is described by the rotation angle $\theta(s)$, function of the curvilinear coordinate $s\in[0,L]$ and the value of the two configurational parameters $s_1$ and $s_2$, associated with the position of the sliding sleeve exits. The rod configuration varies with the change of the sliding sleeves inclinations $\theta_1$ and $\theta_2$ and distance $d$. Resultant force $R$ (inclined at an angle $\beta$) and moments $M_1$ and $M_2$ at the two sliding sleeve exits are also reported. (Right) Curvilinear coordinate $s$ and position of the sliding sleeves' exit at $s_1$ and $s_2=s_1+\ell$, being $\ell$ the rod's length between the two constraints. }
    \label{sist}
    \end{center}
\end{figure}

The planar mechanical behaviour is investigated for  a flexible, unshearable, and inextensible rod constrained at its two terminal parts by two different sliding sleeves, Fig. \ref{sist}. The  position and inclination of each sliding sleeve is controlled.  
The curvilinear coordinate $s$, corresponding to  the arc length because of the inextensibility assumption, is introduced to describe the cross section position along the rod of length $L$, $s\in[0,L]$. The relative position of the two sliding sleeves' exits along the rod  is defined by the two coordinates $s_1$ and $s_2$ ($0\leq s_1 <s_2\leq L$) representing the two  configurational parameters of the system. It follows that the part of rod outside of the sliding sleeve has length $\ell$ given by
\beq\label{elldefinition}
\ell=s_2-s_1.
\eeq

Neglecting rigid-body motion and excluding out-of-plane deformations, the system response is  planar and can be modeled within a Cartesian reference  system $x-y$, centered at the exit of the  sliding sleeve at $s_1$ and with $x-$axis joining the two sliding sleeves' exit points, far from  each other by the distance $d$, so that
\beq\label{coordinates}
x(s_1)=y(s_1)=y(s_2)=0,\qquad x(s_2)=d.
\eeq
The length $\ell$ of the rod between the two constraints is bounded from above by the rod's length $L$ and, due to inextensibility, from below by the distance $d$ between the two sliding sleeve exits  
\beq
d\leq \ell\leq L.
\eeq
Measuring through the rotation field $\theta(s)$ the anti-clockwise angle of the rod's axis with respect to the $x-$axis, the  inclinations $\theta_1$ and $\theta_2$ of the two sliding sleeves introduce the following constraints on  $\theta(s)$
\begin{equation}
\theta(s)=
\left\{
\begin{array}{ll}
\label{constraints0}
\theta_1,\quad s\in[0,s_1],\\ \theta_2,\quad s\in[s_2,L].
\end{array}
\right.
\end{equation}
The constraints (\ref{constraints0}) introduce the possibility that the rotation field $\theta(s)$ is not continuously differentiable at $s_1$ and $s_2$.
Moreover, by considering inextensibility of the rod, the $x(s)-y(s)$ position fields  are constrained to the rotation field $\theta(s)$ through the following differential equations (a prime stands for derivative with respect to $s$)
\beq\label{posrot}
x'(s)=\cos\theta(s),\qquad
y'(s)=\sin\theta(s),
\eeq
and therefore the  two following isoperimetric constraints can be derived from the  coordinates of the two sliding sleeves exits  (\ref{coordinates}) 
\begin{equation}
\label{constraints1}
 \int_{s_1}^{s_2}\cos\theta(s)\,\text{d}s=d,\quad \int_{s_1}^{s_2}\sin\theta(s)\,\text{d}s=0.
\end{equation}

\subsection{Total potential energy and perturbed configuration}\label{sezVAR}

An inextensible and unshearable rod is assumed to  deform only within the $x_1$--$x_2$ plane and to be straight in its undeformed configuration. It is further assumed that the rod is linear elastic, therefore the bending moment $M$ at the curvilinear coordinate $s$ is given by $M(s) =B\theta'(s)$, where $\theta'(s)$ is the rod's curvature and $B$ is the  bending stiffness, considered uniform. 

The   total potential energy $\mathcal{V}$ of the planar elastic system is  defined by the following functional of the rotation field $\theta(s)$ and of the two configurational parameters $s_1$ and $s_2$
\begin{equation}
\begin{split}\label{potentialsling}
\mathcal{V}\left(\theta(s),s_1,s_2\right)=&\frac{B}{2}\int_{s_1}^{s_2}\theta'(s)^2\,\text{d}s-R_x\left(d-\int_{s_1}^{s_2}\cos\theta(s)\,\text{d}s\right)+R_y\int_{s_1}^{s_2}\sin\theta(s)\,\text{d}s\\
&+M_1\left[\theta(s_1)-\theta_1\right]+M_2\left[\theta(s_2)-\theta_2\right],
\end{split}
\end{equation}
where $R_x$, $R_y$, $M_1$, and $M_2$ are the Lagrangian multipliers. $M_1$ and $M_2$  are the moment reactions at the two sliding sleeve exits, enforcing there the controlled sliding sleeves rotations (\ref{constraints0}), while $R_x$ and $R_y$ represent the projections along the $x$ and $y$ axes of the reaction force $R$ at both ends and inclined at the angle $\beta \in[-\pi,\pi]$ (positive when anti-clockwise),
\begin{equation}
\label{Rdef0}
R_x=R\cos\beta,\quad R_y=R\sin\beta,
\end{equation} 
and enforcing the distance between the sliding sleeves through the isoperimetric constraints (\ref{constraints1}). It is highlighted that the expression (\ref{potentialsling})  adopted for the definition of the total potential energy $\mathcal{V}$ is an extended version of the functional usually considered (given for the system under consideration by the elastic bending energy only). Indeed, the expression (\ref{potentialsling}) includes the presence of the null work performed by the Lagrangian multipliers, which  does not affect the total potential energy value but is essential to derive the proper equilibrium equations of  systems subject to isoperimetric constraints.

In order to achieve the equilibrium configuration defined by $s_{1,eq}$, $s_{2,eq}$, and $\theta_{eq}(s)$ and to assess its stability character, a perturbation approach for moving boundaries is now developed.
The configurational parameters $s_1$ and $s_2$ of the sliding sleeve exits  are  considered  through the perturbations $\Delta s_1$ and $\Delta s_2$ superimposed to the respective coordinate at equilibrium
\begin{equation}
\label{perturb0}
s_1=s_{1,eq}+\Delta s_1,\quad
s_2=s_{2,eq}+\Delta s_2.
\end{equation}
Considering that $\ell_{eq}=s_{2,eq}-s_{1,eq}$ and  $\Delta \ell=\Delta s_2-\Delta s_1$, the perturbed external rod's length $\ell$ follows from Eq. (\ref{elldefinition})  as
\begin{equation}
\ell=\ell_{eq}+\ds\Delta \ell.
\end{equation} 

As far as regards the rotation field $\theta(s)$, 
the following perturbed field is considered 
\beq\label{perturba1}
\theta(s+\Delta s_1)=\theta_{eq}(s)+\Delta \theta(s).
\eeq
At this point, it is worth to underline that, beside the choice (\ref{perturba1}), other definitions for the  perturbation in the rotation field can be introduced to analyze the present variational problem with two moving boundaries. For instance,  another option could be the one reported by Elsgolts \cite{elsgolts}
\beq\label{perturba2}
\theta(s)=\theta_{eq}(s)+ \overline{\Delta \theta}(s),
\eeq
thus introducing a perturbation of the equilibrium rotation  $\theta_{eq}(s)$ at each coordinate $s$, while that given by Eq. (\ref{perturba1}) is defined at the perturbed coordinate $s+\Delta s_1$. Due to its cumbersomeness, a further variational approach based on the most general  perturbation in the curvilinear coordinate $s$ (proposed by Gelfand and Fomin \cite{gelfand}) 
is deferred to  Appendix \ref{AppendixA}. It is shown that this approach leads to the  same conclusions  obtained  by considering the perturbation measure $\Delta\theta(s)$, Eq. (\ref{perturba1}).

By comparing Eq. (\ref{perturba1}) with Eq. (\ref{perturba2}), the relation between the two measures of the perturbation in the rotation  is 
\beq\label{legameFDC}
\overline{\Delta \theta}(s)=\theta_{eq}(s-\Delta s_1)- \theta_{eq}(s)+\Delta \theta(s-\Delta s_1).
\eeq
Although the two measures $\Delta \theta(s)$ and $\overline{\Delta \theta}(s)$  of the rotation perturbation  appear equivalent,  it is shown in the next Subsection that only $\Delta \theta(s)$ would allow for describing a  \lq pure-sliding' of the system configuration in a  first order analysis.

\subsection{\lq Pure-sliding'  and the proper measure of rotation perturbation}
\label{puresliding}

\begin{figure}[!h]
\renewcommand{\figurename}{\footnotesize{Fig.}}
    \begin{center}
   \includegraphics[width=0.99\textwidth]{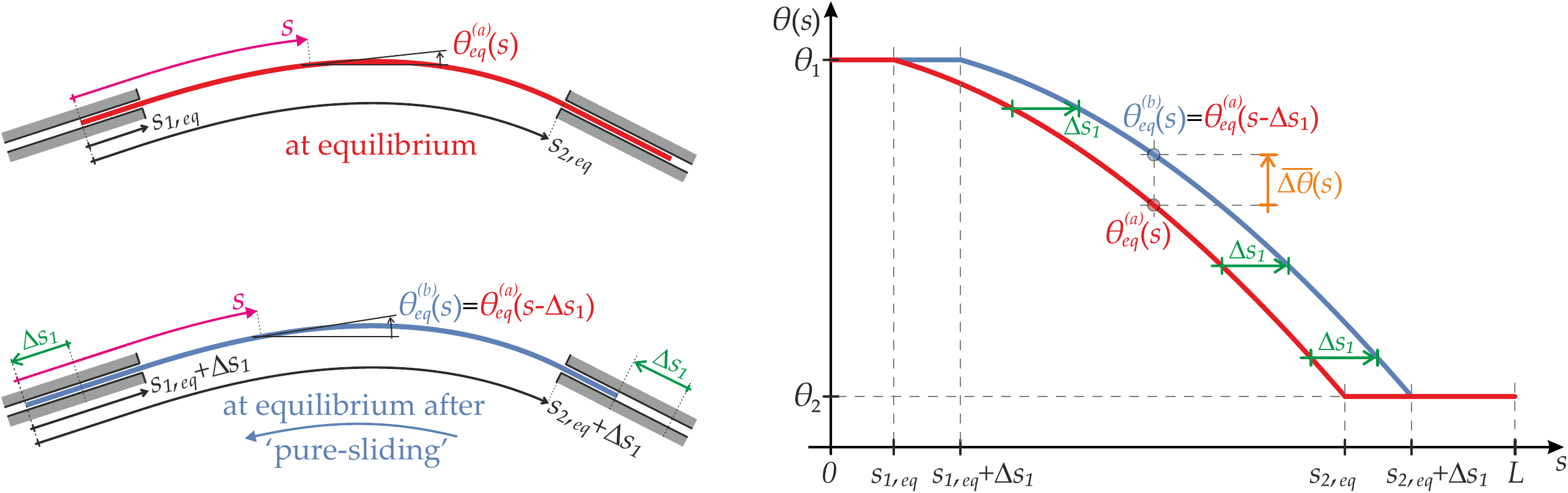}
    \caption{\footnotesize{Neutrality of equilibrium realized through a  \lq pure-sliding' of the elastic rod, made possible by the presence of two sliding sleeves.  (Left) Deformed equilibrium configuration $\theta_{eq}^{(b)}(s)$ (blue) obtained through a \lq pure-sliding' perturbation for a distance $\Delta s_1$ (while $\Delta\theta(s)=0$) of another  equilibrium configuration $\theta_{eq}^{(a)}(s)$  (red). (Right) Equilibrium rotation fields  $\theta_{eq}^{(a)}(s)$  and $\theta_{eq}^{(b)}(s)$  as functions of the curvilinear coordinate $s\in[0,L]$. The non-null variation measure $\overline{\Delta \theta} (s)$ for a \lq pure-sliding' is highlighted.
    }
		}
    \label{fig_shiftfranz}
    \end{center}
\end{figure}

A \lq pure-sliding' perturbation is considered, defined 
as the change in the configuration such that the rod slides in one direction by keeping its portion unconstrained by the sliding sleeves, $s\in[s_1,s_2]$, appearing with the same deformed configuration along a constant  length $\ell$, Fig. \ref{fig_shiftfranz}. With reference to the introduced perturbation quantities, a \lq pure-sliding' perturbation is realized when
\begin{equation}
   \left\{ \begin{array}{ll}
         \Delta s_1=\Delta s_2\neq 0,  \\
         \Delta \theta(s)=0,
    \end{array}
    \right.
    \qquad
    \Longleftrightarrow
  \qquad
  \left\{  \begin{array}{ll}
         \Delta \ell=0,  \\
         \theta(s)=\theta_{eq}(s-\Delta s_1),
    \end{array}
    \right.
    \end{equation}
and, because assumed not explicitly dependent on the curvilinear coordinate $s$ (for instance through a non-constant bending stiffness $B(s)\neq \textup{const}$ or through the presence of an external  load applied at a fixed coordinate $s$), the total potential energy $\mathcal{V}$ is preserved,
   \begin{equation}
\begin{split}
\mathcal{V}\left(\theta(s),s_1,s_2\right)=\mathcal{V}\left(\theta_{eq}(s),s_{1,eq},s_{2,eq}\right).
\end{split}
\end{equation}

The  \lq pure-sliding' perturbations is the key for understanding why  the  measure $\Delta \theta(s)$ is privileged with respect to the measure $\overline{\Delta \theta}(s)$. Indeed, while the former is null, the latter is non-null and given by
\begin{equation}
    \overline{\Delta \theta}(s)=\theta_{eq}(s-\Delta s_1)- \theta_{eq}(s)\neq 0.
\end{equation}
Moreover,  a small generic perturbation around the equilibrium configuration can be considered 
through a small positive parameter $\epsilon$ ruling the modulus of the perturbations $\Delta \theta(s)$, $\Delta s_1$, and $\Delta s_2$,  as \beq
\label{perturb}
s_1=s_{1,eq}+\epsilon\delta s_1,\quad
s_2=s_{2,eq}+\epsilon\delta s_2,\quad
\theta(s+\epsilon \delta s_1)=\theta_{eq}(s)+\epsilon \delta \theta(s),
\eeq
from which it follows
\begin{equation}
\label{perturbell}
\ell=\ell_{eq}+\epsilon \delta \ell,\qquad 
\mbox{where}\qquad\delta \ell=\delta s_2-\delta s_1.
\end{equation}
By assuming the first-order perturbations $\delta\theta(s)$ and $\delta s_1$, the expansion of Eq. (\ref{legameFDC}) for small $\epsilon$
provides the perturbation $\overline{\Delta \theta}(s)$ as the following power series in $\epsilon$
\beq\label{expbarra}
\overline{\Delta \theta}(s)=-\sum_{k=1}^\infty\frac{(-1)^k\,\,\left[k\delta \theta^{[k-1]}(s)-\delta s_1\,\,\theta_{eq}^{[k]}(s)\right]}{k!}\delta s_1^{k-1}\,\epsilon^k,
\eeq
where the superscript $^{[k]}$ defines  the $k$-th derivative (and therefore the superscript $^{[0]}$ stands for no derivative).
Eq. (\ref{expbarra}) further shows that, when a first-order perturbation in $\Delta s_1$ and $\Delta \theta(s)$ is assumed, the \lq pure-sliding' perturbation (simply defined by $\delta\theta(s)=0$ and $\delta\ell=0$)  requires the following infinite-order expansion for $\overline{\Delta \theta}(s)$
\beq
\overline{\Delta \theta}(s)=\sum_{k=1}^\infty\frac{(-1)^k\,\,\theta_{eq}^{[k]}(s)}{k!}\delta s_1^{k}\,\epsilon^k,
\eeq
which discloses that the perturbation $\overline{\Delta \theta}(s)$ could not be considered as a primary perturbation field to investigate the present mechanical system because not including the \lq pure-sliding' perturbation at its first-order expansion. Nevertheless, the measure $\overline{\Delta \theta}(s)$ for the rotation variation is commonly adopted through a small variation $\overline{\delta \theta}(s)$ and leads to  potential energy variations at  different orders coincident with those obtained by considering the small variation $\delta \theta(s)$, as shown in Appendix \ref{AppendixA}.

\subsection{Variational approach}
The planar elastic system with two variable ends is analyzed through a variational approach  by considering the perturbed quantities expressed as in Eq. (\ref{perturb})
with reference to the first-order perturbations $\delta\theta(s)$, $\delta s_1$, and  $\delta s_2=\delta s_1+\delta \ell$.
By considering the perturbed coordinates $s_1$ and $s_2$ and rotation field $\theta(s)$ as given by Eq. (\ref{legameFDC}), the total potential energy $\mathcal{V}$ becomes
\begin{equation}
\begin{split}
\mathcal{V}=&\frac{B}{2}\int_{s_{1,eq}+\epsilon\delta s_1}^{s_{2,eq}+\epsilon\delta s_2}\left[\theta_{eq}'(s-\epsilon\delta s_1)+\epsilon\delta\theta'(s-\epsilon\delta s_1)\right]^2\,\text{d}s\\[2mm]
&-
R_x
\left(d-\int_{s_{1,eq}+\epsilon\delta s_1}^{s_{2,eq}+\epsilon\delta s_2}\cos\left[\theta_{eq}(s-\epsilon\delta s_1)+\epsilon\delta\theta(s-\epsilon\delta s_1)\right]\,\text{d}s\right)\\[2mm]
&+
R_y
\int_{s_{1,eq}+\epsilon\delta s_1}^{s_{2,eq}+\epsilon\delta s_2}\sin\left[\theta_{eq}(s-\epsilon\delta s_1)+\epsilon\delta\theta(s-\epsilon\delta s_1)\right]\,\text{d}s\\[2mm]
&+
M_1
\epsilon \delta\theta(s_{1,eq})+
M_2
\left[\theta_{eq}(s_{2,eq}+\epsilon\delta\ell)+\epsilon\delta\theta(s_{2,eq}+\epsilon\delta\ell)-\theta_2\right],
\end{split}
\end{equation}
which, through a change in the integration variable, can be rewritten as
\begin{equation}
\begin{split}
\mathcal{V}=&\frac{B}{2}\int_{s_{1,eq}}^{s_{2,eq}+\epsilon\delta \ell}\left[\theta_{eq}'(s)+\epsilon\delta\theta'(s)\right]^2\,\text{d}s-
R_x
\left(d-\int_{s_{1,eq}}^{s_{2,eq}+\epsilon\delta \ell}\cos\left[\theta_{eq}(s)+\epsilon\delta\theta(s)\right]\,\text{d}s\right)\\[2mm]
&+
R_y
\int_{s_{1,eq}+\epsilon\delta s_1}^{s_{2,eq}+\epsilon\delta s_2}\sin\left[\theta_{eq}(s)+\epsilon\delta\theta(s)\right]\,\text{d}s+
M_1 \epsilon \delta\theta(s_{1,eq})
+M_2 \left[\theta_{eq}(s_{2,eq}+\epsilon\delta\ell)+\epsilon\delta\theta(s_{2,eq}+\epsilon\delta\ell)-\theta_2\right].
\end{split}
\end{equation}
The last expression, derived under the assumption of  uniform bending stiffness $B(s)=B$,  shows that \beq
\mathcal{V}=\mathcal{V}\left(\theta_{eq}(s),s_{1,eq},s_{2,eq},\delta\theta(s),\delta\ell\right)
\eeq
and therefore that a change in the total potential energy $\mathcal{V}$ is only provided by the perturbation in the rotation field $\delta\theta(s)$ and in the external rod's length $\delta \ell$. Therefore, a \lq pure-sliding' perturbation described by $\delta s_1=\delta s_2\neq 0$ and $\delta\theta(s)=0$ does not vary  the total potential energy $\mathcal{V}$, so  it can be concluded that 
\begin{enumerate}[(i.)]
\item  the coordinate $s_{1,eq}$ plays just the role of a dummy variable and therefore, under the assumption that $s_{1,eq}>0$ and $\ell_{eq}\in(d,L-s_{1,eq})$, the system can be investigated with  reference to the rotation field $\theta_{eq}(s)$ and only one configurational parameter $\ell_{eq}$ (instead of two, $s_{1,eq}$ and $s_{2,eq}=s_{1,eq}+\ell_{eq}$);
\item the planar equilibrium configurations of a uniform rod constrained by two sliding sleeves at its ends are coincident with those corresponding to the case when one of the two sliding sleeves is replaced by a clamp.
\end{enumerate}

In order to obtain the governing equations for the present system and to assess the stability of the equilibrium configurations, the total potential energy $\mathcal{V}$ is expanded as the following power series in the small positive parameter $\epsilon$ 
\begin{equation}
\mathcal{V}\left(\theta(s),s_1,s_2\right)=\mathcal{V}\left(\theta_{eq}(s),\ell_{eq}\right)+
\sum_{k=1}^\infty\dfrac{\epsilon^k}{k!}\delta^k\mathcal{V}\left(\theta_{eq}(s),\ell_{eq},\delta\theta(s),\delta \ell\right),
\end{equation}
where  $\delta^k\mathcal{V}$ represents the $k$-th variation of $\mathcal{V}$, null in the case of a \lq pure-sliding' ($\delta s_1=\delta s_2\neq 0$, $\delta \ell =0$),
\beq
\delta^k\mathcal{V}\left(\theta_{eq}(s),\ell_{eq},\delta\theta(s)=0,\delta \ell=0\right)=0, \qquad \forall\, k\in \mathbb{N}.
\eeq
Interestingly, the boundary conditions (\ref{constraints0}) evaluated at $s_1$ and at $s_2$ and expanded for small $\epsilon$ imply the following compatibility conditions for the perturbation $\delta\theta(s)$ 
\beq\label{compatibility}\barr{lll}
\delta\theta(s_{1,eq})=0,\\
\delta\theta^{[k]}(s_{2,eq})=-\dfrac{\theta_{eq}^{[k+1]}(s_{2,eq})\,\delta \ell}{k+1},\quad k \in \mathbb{N}_1,
\earr
\eeq
while the first-order expansion of the isoperimetric constraints (\ref{constraints1}) imply
\begin{equation}\label{isoperimetricderived}
    \cos\theta_{eq}(s_{2,eq}) \delta \ell
- \int_{s_{1,eq}}^{s_{2,eq}}\sin\theta_{eq}(s)\delta \theta(s)\,\text{d}s=0,
\qquad
\sin\theta_{eq}(s_{2,eq}) \delta \ell
+ \int_{s_{1,eq}}^{s_{2,eq}}\cos\theta_{eq}(s)\delta \theta(s)\,\text{d}s=0.
\end{equation}
It is noted that the perturbations evaluated at $s_{1,eq}$ and $s_{2,eq}$ values are intended as the value of the perturbation by approaching the ends from outside the constrained region, namely, $s_{1,eq}\rightarrow s_{1,eq}^+$ and $s_{2,eq}\rightarrow s_{2,eq}^-$, because  the rotation field $\theta(s)$ is not continuously differentiable at these two points from the constraints (\ref{constraints0}).

After integration by parts, exploiting the boundary conditions (\ref{constraints0}) and the compatibility constraints (\ref{compatibility}) on the perturbations $\delta\theta(s)$ and $\delta\ell$,  the first variation $\delta\mathcal{V}$ is given by
\begin{equation}
\label{1varsling}
\begin{array}{lll}
\delta\mathcal{V}=
&
\ds-\int_{s_{1,eq}}^{s_{2,eq}}\left[B\theta_{eq}''(s)+R_x\sin\theta_{eq}(s)-R_y\cos\theta_{eq}(s)\right]\delta\theta(s)\,\text{d}s
 \\[2mm]
 &\ds-\left\{\frac{B \left[\theta_{eq}'(s_{2,eq})\right]^2}{2}-R_x\cos\theta_{eq}(s_{2,eq})-R_y\sin\theta_{eq}(s_{2,eq})\right\}\delta \ell.
\end{array}
\end{equation}

The annihilation of the first variation $\delta\mathcal{V}$ for every perturbations $\delta\theta(s)$ and $\delta\ell$
leads to the elastica equation
\beq\label{elastica}
B\theta_{eq}''(s)+R_x\sin\theta_{eq}(s)-R_y\cos\theta_{eq}(s)=0,\qquad
s\in(s_{1,eq},s_{2,eq}),
\eeq
and to the interface condition at the sliding sleeve at $s=s_{2,eq}$
\beq
\label{interface}
\frac{B \left[\theta_{eq}'(s_{2,eq})\right]^2}{2}-R_x\cos\theta_{eq}(s_{2,eq})-R_y\sin\theta_{eq}(s_{2,eq})=0.
\eeq
Manipulation of the elastica equation (\ref{elastica}) and the interface condition (\ref{interface}) provides
\beq\label{invaria}
\frac{B \left[\theta_{eq}'(s)\right]^2}{2}-R_x\cos\theta_{eq}(s)-R_y\sin\theta_{eq}(s)=0, \qquad
s\in\left(s_{1,eq},s_{2,eq}\right).
\eeq

Recalling the definition  (\ref{Rdef0}) for the reaction components $R_x$ and $R_y$, the last equation yields
\beq\label{invaria2}
R \cos[\theta_{eq}(s)-\beta]=\frac{B \left[\theta_{eq}'(s)\right]^2}{2}\geq0, \qquad
s\in\left(s_{1,eq},s_{2,eq}\right),
\eeq
which in turn implies
\beq\label{invaria3}
-\dfrac{\pi}{2}\leq\theta_{eq}(s)-\beta\leq\dfrac{\pi}{2}, \qquad
s\in\left(s_{1,eq},s_{2,eq}\right).
\eeq
Inequality (\ref{invaria3}) provides the following  necessary condition for the existence of the equilibrium configuration
 \begin{equation}\label{bound0}
     \left|\max_{s\in\left(s_{1,eq},s_{2,eq}\right)}\left\{\theta(s)\right\}
     -
     \min_{s\in\left(s_{1,eq},s_{2,eq}\right)}\left\{\theta(s)\right\}\right|\leq\pi.
 \end{equation}
  In the next Sect. it is shown that the equilibrium field $\theta(s)$ may display a number $m\geq 0$ of inflection points at the curvilinear coordinate $\widehat{s}_j$ ($j=1,...,m$), therefore defined by a  null derivative of the rotation at these points, $\theta'(\widehat{s}_j)=0$. In the case of elasticae with no inflection points ($m=0$), 
 the necessary condition  (\ref{bound0}) is equivalent to
 \begin{equation}\label{bound}
     \left|\theta_1-\theta_2\right|\leq\pi.
 \end{equation}
 In the case of elasticae with inflection points  ($m\geq 1$), Eq. (\ref{bound}) still represents a necessary condition, although less restrictive than Eq. (\ref{bound0}).
Moreover, in the case of elasticae with inflection points  ($m\geq 1$), Eq. (\ref{invaria2}) shows that at equilibrium  the reaction force $R$ is  orthogonal to the rod's tangent
 at each inflection point, 
\beq\label{ortho}
\left|\theta(\widehat{s}_j)-\beta\right|=\frac{\pi}{2},\qquad j=1,...,m.
\eeq

Considering equation (\ref{invaria}) and the compatibility conditions (\ref{compatibility}), the second variation $\delta^2 \mathcal{V}$ can be expressed by
\begin{equation}
\begin{array}{lll}\label{secondvarsling}
\delta^2 \mathcal{V}=&\ds B\int_{s_{1,eq}}^{s_{2,eq}}\left\{\,\left[\delta \theta'(s)\right]^2-\frac{ \left[\theta_{eq}'(s)\right]^2}{2}\left[\delta \theta(s)\right]^2 \right\}\,\text{d}s-B\theta_{eq}'(s_{2,eq})\theta_{eq}''(s_{2,eq})\delta \ell^2,
\end{array}
\end{equation}
while the  third variation $\delta^3 \mathcal{V}$ by
\begin{equation}\label{thirdVarV}
\begin{array}{lll}
\delta^3 \mathcal{V}=&\ds-B\int_{s_{1,eq}}^{s_{2,eq}}\,\theta_{eq}''(s)\,\left[\delta \theta(s)\right]^3\,\text{d}s-\frac{B}{2}\left\{\frac{\left[\theta_{eq}''(s_{2,eq})\right]^2}{2}+\left[\theta_{eq}'(s_{2,eq})\right]^4\right\}\delta \ell^3.
\end{array}
\end{equation}

From the stability point of view under conservative conditions, due to the presence of possible \lq pure-sliding' perturbations, the planar equilibrium configurations of the rod when constrained by two sliding sleeves can be neutral or unstable, respectively corresponding to stable and unstable for the same system where a sliding sleeve is replaced by a clamp. However, the inherent neutrality of the equilibrium of the system with two sliding sleeves  can be never observed in practical realizations because of  unavoidable presence of (air drag and/or friction) dissipation mechanisms. Therefore, 
\begin{center}
\emph{
from the practical point of view, the planar mechanics of a uniform elastic rod\\ constrained by two sliding sleeves is equivalent to that of\\ the same system with one of the two sliding sleeves  replaced by a clamp.}
\end{center}
It is worth to highlight that the latter statement holds only for systems characterized by a total potential energy $\mathcal{V}$ not  explicitly dependent on the curvilinear coordinate $s$ (therefore, for instance, the application of forces at fixed coordinate $s$ and  a non-constant bending stiffness $B(s)\neq \textup{const}$ are excluded).

The equilibrium configurations and the stability criterion for the present planar system are addressed in the next two Sections. Since only the configurations at equilibrium are considered henceforth, the subscript $(\cdot)_{eq}$ is removed to simplify the notation, namely
\beq
\theta_{eq}(s)\rightarrow\theta(s),\qquad 
s_{1,eq}\rightarrow
s_1,
\qquad
s_{2,eq}\rightarrow
s_2,
\qquad
 \ell_{eq}\rightarrow
\ell.
\eeq

\section{Equilibrium configurations}
\label{eqconfs}
\begin{figure}[!h]
\renewcommand{\figurename}{\footnotesize{Fig.}}
    \begin{center}
  \includegraphics[width=0.85\textwidth]{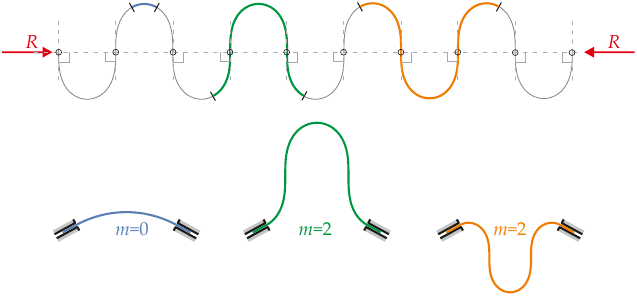}
\caption{\footnotesize{(Above) Among the infinite planar equilibrium  configurations for $\theta_1=-\theta_2=\pi/8$, the non-inflectional (blue) and two inflectional elasticae with $m=2$ (green and orange) are highlighted along the  \lq inflectional mother curve' (grey curve).
The \lq inflectional mother curve' is associated to equilibrium configurations with $\eta=1/\sqrt{2}$, for which  the reaction force $R$ is orthogonal to the elastica at each inflection point (circles).  
(Bottom) The three equilibrium configurations highlighted along the \lq inflectional mother curve'  for $\theta_1=-\theta_2=\pi/8$ reported for the same sliding sleeves' distance $d$, showing  different values of the corresponding external lengths $\ell$.
    }
		}
    \label{fig_mothercurve}
    \end{center}
\end{figure}

The  closed-form solution  for the elastica subject to Dirichlet boundary conditions and isoperimetric constraints can be obtained by distinguishing  the two fundamental cases of  absence and presence of inflection points, whose number is denoted by $m$, and therefore corresponding to a non-inflectional and an inflectional configuration, respectively.  While the curvature maintains the same sign for $s\in[s_1,s_2]$ in the former case, a sign change occurs at $m$ points within the same set in the latter case and, as a result, the corresponding solution is more involved. The solution of the equilibrium equation (\ref{elastica})  has been  presented  in \cite{cazzolli}, showing that the sliding sleeve reaction force  $R$ is given by
\begin{equation}
\label{Rdef}
\dfrac{R\ell^2}{B}= \left\{
\begin{array}{lll}
\xi^2\left[\mathcal{F}(\chi_{2},\,\xi)-\mathcal{F}(\chi_{1},\,\xi)\right]^2,\quad\quad&m=0,\\[3mm]
 \left[\mathcal{F}(\omega_{2},\,\eta)-\mathcal{F}(\omega_1,\,\eta)\right]^2,\quad\quad &m\neq 0,
\end{array}
\right.
\end{equation} 
while the rotation field $\theta(s)$ for the part of rod outside of the sliding sleeves, $s\in[s_1,s_2]$, by 
\begin{equation}\label{rotasol}
\theta(s)= \left\{
\begin{array}{lll}
\beta+2\,\textup{am}\left(\dfrac{s-s_1}{\ell}\left(\mathcal{F}(\chi_{2},\,\xi)-\mathcal{F}(\chi_{1},\,\xi)\right)+\mathcal{F}(\chi_{1},\,\xi),\,\xi\right), \qquad\qquad & m=0,\\[3mm]
\beta+2\,\arcsin\left[\eta\,\textup{sn}\left(\dfrac{s-s_1}{\ell}\left(\mathcal{F}(\omega_{2},\,\eta)-\mathcal{F}(\omega_1,\,\eta)\right)+\mathcal{F}(\omega_1,\,\eta),\,\eta\right)\right],\qquad\qquad & m\neq 0.
\end{array}
\right.
\end{equation}
In Eqs. (\ref{Rdef}) and (\ref{rotasol}),  $\mathcal{F}$ is the \textit{Jacobi's incomplete elliptic integral of the first kind}, $\textup{am}$ is the \textit{Jacobi's amplitude function} and $\textup{sn}$  is the \textit{Jacobi's sine amplitude function},
\begin{equation}
 \mathcal{F}\left(\sigma,\,\varphi\right)=\int_{0}^{\sigma}\frac{\textup{d}\phi}{\sqrt{1-\varphi^2\sin^2{\phi}}},
 \qquad
 \sigma=\textup{am}\bigg( \mathcal{F}\left(\sigma,\,\varphi\right),\,\varphi\bigg),
 \qquad
 \textup{sn}(u,\,\varphi)=\sin{\left(\textup{am}(u,\,\varphi)\right)},
\end{equation}
and 
\begin{equation}
\begin{array}{ccc}
\label{parapara}
\chi_1=\dfrac{\theta_1-\beta}{2},\quad
\chi_{2}=\dfrac{\theta_2-\beta}{2},\quad \omega_1=\arcsin{\left(\dfrac{1}{\eta}\sin{\dfrac{\theta_{1}-\beta}{2}}\right)},\\
\omega_{2}=(-1)^m\,\arcsin{\left(\dfrac{1}{\eta}\sin{\dfrac{\theta_{2}-\beta}{2}}\right)}+(-1)^p\,m\pi,
\end{array}
\end{equation}
where the Boolean parameter $p$ is introduced to define the solution with positive (if $p=0$) or negative (if $p=1$) sign of the curvature at the initial end $s=s_1$. Moreover, defining $\widehat{\theta}=\theta(\widehat{s})$ as the rotation value at the  inflection point with smallest coordinate value $\widehat{s}=\widehat{s}_1$, $\theta'(\widehat{s})=0$, the parameter $\eta$ is given by
\begin{equation}
\eta=\sin\frac{\widehat{\theta}-\beta}{2}.
\end{equation}
Imposing condition (\ref{invaria}) on the generic solution (\ref{rotasol})  in the absence of inflection points ($ m=0$) reveals the value of the parameter $\xi$ 
\begin{equation}
\frac{B}{\ell^2}\left[\mathcal{F}(\chi_{2},\,\xi)-\mathcal{F}(\chi_{1},\,\xi)\right]^2(2-\xi^2)=0\quad \Rightarrow \quad 
\xi=\sqrt{2},
\end{equation}
and in the presence of inflection points ($ m\neq0$) that of $\eta$
\begin{equation}
\frac{B}{\ell^2}\left[\mathcal{F}(\omega_{2},\,\eta)-\mathcal{F}(\omega_1,\,\eta)\right]^2(2\eta^2-1)=0
\quad \Rightarrow \quad 
\eta=\dfrac{1}{\sqrt{2}},
\end{equation}
the latter further  confirming the orthogonality  between the direction of the resultant $R$ and the rod's tangent at the inflection coordinate $\widehat{s}$, Eq. (\ref{ortho}).
 It also follows that all the equilibrium configurations with inflection points are different portions  of the same \lq inflectional mother curve' \cite{born, cazzolli, frish, love} characterized by property (\ref{ortho}), Fig. \ref{fig_mothercurve}. According to previous investigation \cite{cazzolli}, the equilibrium configuration with no inflection point $m=0$, therefore related to the parameter $\xi=\sqrt{2}$, represent portions being \lq extracted' from the inflectional mother curve, related to the parameter $\eta=1/\sqrt{2}$. 
 
Considering the rotation field $\theta(s)$, Eq. (\ref{rotasol}), in the integration of the differential relations (\ref{posrot}), 
the closed-form expressions for  the position field $x(s)$ and $y(s)$ can be obtained  as 
\begin{equation}
\label{x1x2}
	\begin{bmatrix}
	x(s) \\ y(s)
	\end{bmatrix}
	=\ell
	\begin{bmatrix}
	\cos{\beta} & -\sin{\beta}\\
	\sin{\beta} & \cos{\beta}
	\end{bmatrix}\cdot
	\begin{bmatrix}
	\mathbb{A} \left(s\right)\\ \mathbb{B}(s)
	\end{bmatrix}\qquad s\in[s_1,s_2],
\end{equation}
where, in the absence of inflection points ($m=0$), $\mathbb{A}(s)$ and $\mathbb{B}(s)$ are given by 
\begin{equation}
\label{Afun}
\begin{array}{ll}
\mathbb{A}(s)= 
\dfrac{\mathscr{E}\left[\dfrac{s-s_1}{\ell}\left(\mathcal{F}\left(\chi_{2},\,\sqrt{2}\right)-\mathcal{F}\left(\chi_{1},\,\sqrt{2}\right)\right)+\mathcal{F}\left(\chi_{1},\,\sqrt{2}\right),\,\sqrt{2}\right]
-
\mathscr{E}\left[\mathcal{F}\left(\chi_1,\,\sqrt{2}\right),\,\sqrt{2}\right]}{\mathcal{F}\left(\chi_{2},\,\sqrt{2}\right)-\mathcal{F}\left(\chi_1,\,\sqrt{2}\right)},\\[6mm]
\mathbb{B}(s)=
\dfrac{\textup{dn}\left[\mathcal{F}\left(\chi_1,\,\sqrt{2}\right),\,\sqrt{2}\right]-\textup{dn}\left[\dfrac{s-s_1}{\ell}\left(\mathcal{F}\left(\chi_{2},\,\sqrt{2}\right)-\mathcal{F}\left(\chi_{1},\,\sqrt{2}\right)\right)+\mathcal{F}\left(\chi_{1},\,\sqrt{2}\right),\,\sqrt{2}\right]}{\mathcal{F}\left(\chi_{2},\,\sqrt{2}\right)-\mathcal{F}\left(\chi_1,\,\sqrt{2}\right)},
\end{array}
\end{equation}
while in the presence of inflection points ($m\neq0$) by
\begin{equation}
\label{Bfun}
\begin{array}{lll}
\mathbb{A}(s)= \resizebox{0.88\textwidth}{!}{$
2\dfrac{\mathscr{E}\left[\dfrac{s-s_1}{\ell}\left(\mathcal{F}\left(\omega_{2},\,\dfrac{1}{\sqrt{2}}\right)-\mathcal{F}\left(\omega_1,\,\dfrac{1}{\sqrt{2}}\right)\right)+\mathcal{F}\left(\omega_1,\,\dfrac{1}{\sqrt{2}}\right),\,\dfrac{1}{\sqrt{2}}\right]
-\mathscr{E}\left[\mathcal{F}\left(\omega_1,\,\dfrac{1}{\sqrt{2}}\right),\,\dfrac{1}{\sqrt{2}}\right]}{\mathcal{F}\left(\omega_{2},\,\dfrac{1}{\sqrt{2}}\right)-\mathcal{F}\left(\omega_1,\,\dfrac{1}{\sqrt{2}}\right)}\,
-\dfrac{s-s_1}{\ell}$},\\[8mm]
\mathbb{B}(s)=\resizebox{0.88\textwidth}{!}{$
\sqrt{2}\,\dfrac{\textup{cn}\left[\mathcal{F}\left(\omega_1,\,\dfrac{1}{\sqrt{2}}\right),\,\dfrac{1}{\sqrt{2}}\right]-\textup{cn}\left[\dfrac{s-s_1}{\ell}\left(\mathcal{F}\left(\omega_{2},\,\dfrac{1}{\sqrt{2}}\right)-\mathcal{F}\left(\omega_1,\,\dfrac{1}{\sqrt{2}}\right)\right)+\mathcal{F}\left(\omega_1,\,\dfrac{1}{\sqrt{2}}\right),\,\dfrac{1}{\sqrt{2}}\right]}{\mathcal{F}\left(\omega_{2},\,\dfrac{1}{\sqrt{2}}\right)-\mathcal{F}\left(\omega_1,\,\dfrac{1}{\sqrt{2}}\right)}.$}
\end{array}
\end{equation}

In the previous equations  the function $\textup{cn}$ is the \textit{Jacobi's cosine amplitude function},
$\mathscr{E}$ is the  \textit{Jacobi's epsilon function}, and $ \textup{dn}$ is the \textit{Jacobi's elliptic function}
\begin{equation}
 \textup{cn}(u,\,\varphi)=\cos{\left(\textup{am}(u,\,\varphi)\right)},\qquad
 \mathscr{E}\left(\sigma,\,\varphi\right)=E(\textup{am}(\sigma,\,\varphi),\,\varphi),\qquad
  \textup{dn}(u,\,\varphi)=\sqrt{1-\varphi^2\,\textup{sn}^2(u,\,\varphi)},
\end{equation}
while $E$ is the \textit{Jacobi's incomplete elliptic integral of
the second kind}
\begin{equation}
 E\left(\sigma,\,\varphi\right)=\int_{0}^{\sigma}\sqrt{1-\varphi^2\sin^2{\phi}}\,\textup{d}\phi.
\end{equation}

Finally, the isoperimetric constraints (\ref{constraints1}) provide  the following weakly-coupled nonlinear algebraic system in the unknown reaction inclination $\beta$ and rod's length $\ell$ for given sliding sleeves' rotations $\theta_1$ and $\theta_2$, and  distance $d$ 
\begin{equation}
\label{solselastica}
		\ell\begin{bmatrix}
	\cos{\beta} & -\sin{\beta}\\
	\sin{\beta} & \cos{\beta}
	\end{bmatrix}\cdot
	\begin{bmatrix}
	\mathsf{A}(\beta)\\ \mathsf{B}(\beta)
	\end{bmatrix}=
	\begin{bmatrix}
	d \\ 0
	\end{bmatrix}
	,
\end{equation}
where $\mathsf{A}(\beta)=\mathbb{A} (s_1+\ell)$ and $\mathsf{B}(\beta)=\mathbb{B}(s_1+\ell)$, namely
\begin{equation}
\label{Afun2}
\mathsf{A}(\beta)=\left\{
\begin{array}{ll}
\dfrac{E\left(\chi_{2},\,\sqrt{2}\right)
-E\left(\chi_1,\,\sqrt{2}\right)
}{
\mathcal{F}\left(\chi_{2},\,\sqrt{2}\right)-\mathcal{F}\left(\chi_1,\,\sqrt{2}\right)},\\[6mm]
2\dfrac{E\left(\omega_{2},\,\dfrac{1}{\sqrt{2}}\right)
-E\left(\omega_1,\,\dfrac{1}{\sqrt{2}}\right)}{\mathcal{F}\left(\omega_{2},\,\dfrac{1}{\sqrt{2}}\right)-\mathcal{F}\left(\omega_1,\,\dfrac{1}{\sqrt{2}}\right)},
\end{array}\right.
\quad
\mathsf{B}(\beta)= \left\{
\begin{array}{lll}
\dfrac{\sqrt{\cos 2\chi_1}-\sqrt{\cos 2\chi_2}}{\mathcal{F}\left(\chi_{2},\,\sqrt{2}\right)-\mathcal{F}\left(\chi_1,\,\sqrt{2}\right)}, &\quad m=0,\\[8mm]
\dfrac{\sqrt{2}\left(\cos\omega_1-\cos\omega_{2}\right)}{\mathcal{F}\left(\omega_{2},\,\dfrac{1}{\sqrt{2}}\right)-\mathcal{F}\left(\omega_1,\,\dfrac{1}{\sqrt{2}}\right)},&\quad m\neq 0.
\end{array}
\right.
\end{equation}
Solving the nonlinear system (\ref{solselastica}) provides one  or more pairs of the unknown parameters $\beta$ and $\ell$ corresponding to planar equilibrium  configurations with $m$ inflection points.
The weak-coupling in the nonlinear system (\ref{solselastica}) allows for computing the value of $\beta$  from
\beq
\mathsf{A}(\beta)\sin{\beta} +\mathsf{B}(\beta) \cos{\beta}=0,
\eeq
which defines a set of infinite homothetic equilibrium configurations characterized by the following  constant ratio for $\ell/d$,
\beq
\frac{\ell}{d}=\dfrac{\cos{\beta}}{\mathsf{A}(\beta)}\geq 1.
\eeq
This implies that
the planar equilibrium configurations have  shape governed only by the parameter $\beta$ (which in turn depends on the sliding sleeves' inclination, $\theta_1$ and $\theta_2$) and is unaffected by the distance parameter $d$, which has the only effect to scale the size of the deformed configuration (as long as $\ell<L$).

The conditions to assess the stability of the non-unique equilibrium configurations are derived in the following Section.  It is anticipated that no more than one  stable configuration exists for each pair of inclinations $\theta_1$ and $\theta_2$, and that the stable configuration is characterized by none ($m=0$) or  one inflection ($m=1$) point only.

\section{Stability criterion for systems subject to isoperimetric constraints and with variable-length  ($\delta \ell\neq 0$)}

An equilibrium configuration is stable whenever the corresponding second variation $\delta^2\mathcal{V}$ in the total potential energy is  positive for every  perturbations $\delta\theta(s)$ and $\delta\ell$ that are compatible, namely, consistent with the whole set of imposed constraints. The equilibrium stability for systems with variable-domain  has so far only been  assessed in the absence of isoperimetric constraints \cite{majidi}.  On the other hand, Bolza in 1902 \cite{bolza} established the stability criterion for systems with  non-variable one-dimensional domains subject to isoperimetric constraints. 

A criterion for assessing the stability of  variable-domain systems under isoperimetric constraints  is introduced for the first time, by extending the formulation by Bolza \cite{bolza} to variable domains. By excluding  \lq pure-sliding' perturbations (Sect. \ref{sezVAR}),  the stability of the planar system with two sliding sleeves is equivalent to that with one of the two constraints replaced by a clamp. This simplified system is here addressed by considering fixed $s_1$ and varying  only $s_2$, namely $\delta s_1=0$ and $\delta s_2=\delta \ell$. The following generic expression for the  second variation $\delta^2 \mathcal{V}$, to be later reduced to that  relevant to the present problem, Eq. (\ref{secondvarsling}),  is considered 
\begin{equation}\label{var2bolza0}
\delta^2 \mathcal{V}=\int_{s_1}^{s_2}\left(H_1\, \delta\theta'^2+H_2 \,\delta\theta^2\right) \text{d}s+\mathscr{F}(s_2)\delta \ell^2,
\end{equation}
where  $H_1(s)>0$ (denoting the  so-called \lq Legendre's condition', necessary for the stability of the system) and $H_2(s)$ are given functions of the spatial variable $s$ ranging within the fixed set $[s_1,s_2]$, while $\mathscr{F}(s_2)$ is a given function of $s_2$. The system is considered to be subject to a boundary condition at the fixed end $s_1$ and to a compatibility condition at the moving end $s_2$, providing the following constraints for the  perturbation field $\delta\theta(s)$ and the length perturbation  $\delta \ell\neq 0$ at these two coordinates
\begin{equation}
\label{BCwvarend_bc}
\delta\theta(s_1)=0,\qquad \delta\theta(s_2)=W(s_2)\delta \ell,
\end{equation}
where $W(s_2)$ is a given function of $s_2$, and also subject to $N$ isoperimetric constraints on the perturbations, expressed through given functions $T_i(s)$ and  scalars $f_i(s_2)$ ($i=1,..., N$) by
\begin{equation}
\label{BCwvarend_iso}
\int_{s_1}^{s_2}\, T_i \delta\theta \text{d}\sigma=f_i(s_2) \delta \ell, \quad \,i=1,..., N.
\end{equation}

Introducing the differential operator $\Psi$  as
\begin{equation}
\label{psioper}
\Psi(\delta\theta)=-(H_1 \,\delta\theta')'+H_2 \,\delta\theta,
\end{equation}
and the  vector of independent functions $\textbf{u}(s)=\{u(s),v_1(s),\dots,v_N(s)\}$ as the solution of the following  differential system
\begin{equation}
\label{functs}
\left\{
\begin{array}{lll}
\Psi(u(s))=0,& u(s_1)=0,& u'(s_1)=1,\\
\Psi(v_1(s))=T_1(s), & v_1(s_1)=0,& v_1'(s_1)=0,\\
\vdots
& \vdots
& \vdots\\
\Psi(v_N(s))=T_N(s),& v_N(s_1)=0,& v_N'(s_1)=0,\\
\end{array}
\right.
\end{equation}
the generic compatible perturbation  $\delta\theta(s)$ can be expressed as the following combination of continuous functions 
\begin{equation}\label{womega}
\delta\theta=\textbf{p}\cdot \textbf{u}= p(s) u(s) + q_1(s) v_1(s)+...+ q_N(s) v_N(s),
\end{equation}
where $\textbf{p}=\{p(s),q_1(s),\dots,q_N(s)\}$ is a vector of arbitrary functions admitting first and second derivatives, while the symbol $\cdot$ represents the scalar product. 

From Eq. (\ref{psioper}) the following relation  can be obtained
\begin{equation}\label{prop11}
\delta\theta\,\Psi(\delta\theta)= \delta\theta \left[H_2\, \textbf{p}\cdot \textbf{u}-\left(H_1(\textbf{p}\cdot\textbf{u})'\right)' \right],
\end{equation}
which, after several passages, can be rewritten as 
\begin{equation}
\label{omegapsi1}
\delta\theta \,\Psi(\delta\theta)= \delta\theta \,\textbf{p}\cdot \Psi(\textbf{u})-\left[\delta\theta\,H_1 \textbf{p}'\cdot \textbf{u}\right]'+H_1(\textbf{p}'\cdot \textbf{u})^2+H_1\left[ (\textbf{p}'\cdot \textbf{u})(\textbf{p}\cdot \textbf{u}') -(\textbf{p}'\cdot \textbf{u}')(\textbf{p}\cdot \textbf{u})\right].
\end{equation}
Considering that $\textbf{u}(s_1)=\textbf{0}$, by definition of the system (\ref{functs}), and that the following   properties hold
\begin{equation}
\label{tensprop}
\textbf{a}\otimes\textbf{b}:\textbf{c}\otimes\textbf{d}=(\textbf{a}\cdot \textbf{c})(\textbf{b}\cdot \textbf{d}),\qquad \textbf{A}:\textbf{B}=\tr{\left[\textbf{A}\cdot \textbf{B}^\intercal\right]},\qquad 
\textbf{a}\otimes\textbf{b}=\left(\textbf{b}\otimes\textbf{a}\right)^\intercal,
\end{equation}
where the symbols \lq$\otimes$', \lq$:$', \lq$\tr$' and \lq$^\intercal$' respectively denote the outer product, the double scalar product, the trace operator and the transpose operator, Eq. (\ref{omegapsi1}) can be rewritten as (details are reported in Appendix \ref{mathdet})
\begin{equation}
\label{omegapsi3}
\begin{split}
\delta\theta\,\Psi(\delta\theta)&= H_1(\textbf{p}'\cdot \textbf{u})^2+\left[\textbf{p}\otimes\textbf{p}:\int_{s_1}^{s}\Psi(\textbf{u})\otimes \textbf{u}\,\text{d}\sigma-\delta\theta\,H_1 (\textbf{p}'\cdot \textbf{u})\right]'-2\,\textbf{p}\otimes\textbf{p}':\int_{s_1}^{s}\Psi(\textbf{u})\otimes \textbf{u}\,\text{d}\sigma.
\end{split}
\end{equation}
Exploiting the  identity derived from Eq. (\ref{prop11}) and by integration by parts
\begin{equation}
\label{newint0}
\begin{split}
H_1\,\delta\theta'^2+H_2\,\delta\theta^2=\delta\theta\,\Psi(\delta\theta)+(H_1\,\delta\theta\, \delta\theta')',
\end{split}
\end{equation}
the following relation finally holds for an arbitrary number $N$ of isoperimetric constraints
\begin{equation}
\label{newint}
\begin{split}
H_1\,\delta\theta'^2+H_2\,\delta\theta^2= H_1(\textbf{p}'\cdot \textbf{u})^2+\left[\textbf{p}\otimes\textbf{p}:\int_{s_1}^{s}\Psi(\textbf{u})\otimes \textbf{u}\,\text{d}\sigma+\delta\theta\,H_1 (\textbf{p}\cdot \textbf{u}')\right]'-2\,\textbf{p}\otimes\textbf{p}':\int_{s_1}^{s}\Psi(\textbf{u})\otimes \textbf{u}\,\text{d}\sigma.
\end{split}
\end{equation}

By restricting attention to the case of two isoperimetric constraints $N=2$, the vectors $\textbf{p}(s)$ and $\textbf{u}(s)$ reduce to
\begin{equation}
\textbf{p}(s)=\left[
\begin{array}{ccc}
p(s)\\
q_1(s)\\
q_2(s)
\end{array}
\right],
\qquad
\textbf{u}(s)=\left[
\begin{array}{ccc}
u(s)\\
v_1(s)\\
v_2(s)
\end{array}
\right],
\end{equation}
and the following identity holds
\begin{equation}
\int_{s_1}^{s}\Psi(\textbf{u}(\sigma))\otimes \textbf{u}(\sigma)\,\text{d}\sigma=\left[
\begin{array}{ccc}
0 & 0 & 0\\
m_1(s) & n_1(s) & o_1(s)\\
m_2(s) & n_2(s) & o_2(s)
\end{array}
\right],
\end{equation}
where 
\begin{equation}
m_i(s)=\int_{s_1}^{s} u(\sigma)\, T_i(\sigma) \text{d}\sigma,\qquad 
n_i(s)=\int_{s_1}^{s} v_1(\sigma)\, T_i(\sigma) \text{d}\sigma,\qquad 
o_i(s)=\int_{s_1}^{s} v_2(\sigma)\, T_i(\sigma) \text{d}\sigma,
\end{equation}
and therefore the last term on the right hand side of Eq. (\ref{newint}) simplifies as
\begin{equation}\label{pdiadepprimo}
-2\,\textbf{p}\otimes\textbf{p}':\int_{s_1}^{s}\Psi(\textbf{u})\otimes \textbf{u}\,\text{d}\sigma=-2q_1(p' m_1+q_1'n_1+q_2'o_1)-2q_2(p' m_2+q_1'n_2+q_2'o_2).
\end{equation}
Following Bolza \cite{bolza}, the following two constraints for the first derivative of $p(s)$, $q_1(s)$, and $q_2(s)$ are considered without loss of generality 
\begin{equation}
p' m_i+q_1'n_i+q_2'o_i=0,\qquad i=1,2.
\label{eq2}
\end{equation}
The latter constraints lead to the following identities
\begin{equation}
(p m_i+q_1 n_i+q_2 o_i)'=T_i (p u+q_1 v_1+q_2 v_2)=T_i \delta\theta,\qquad i=1,2,
\end{equation}
which can be integrated to provide
\begin{equation}
    \label{eq23}
p(s) m_i(s)+q_1(s) n_i(s)+q_2(s) o_i(s)=\int_{s_1}^{s} \,T_i\delta\theta\,\text{d}\sigma,\qquad i=1,2,
\end{equation}
by  recalling that $m_i(s_1)=n_i(s_1)=o_i(s_1)=0$. It follows that  Eq. (\ref{womega}) together with the two constraints (\ref{eq23}) provide a linear  system in  $\textbf{p}$ expressed by
\begin{equation}
\label{systemm}
\mathscr{M}(s)
\textbf{p}(s)=\textbf{z}(s),
\end{equation}
where the square matrix $\mathscr{M}(s)$ and the vector $\textbf{z}(s)$ are given by 
\begin{equation}
\label{matrix}
\mathscr{M}(s)=\left[
\begin{array}{ccc}
m_1(s) & n_1(s) & o_1(s)\\
m_2(s) & n_2(s) & o_2(s)\\
u(s) & v_1(s) & v_2(s)
\end{array}
\right],\qquad
\textbf{z}(s)=\left[
\begin{array}{ccc}
\ds \int_{s_1}^{s}\,T_1(\sigma)\delta\theta(\sigma)\,\text{d}\sigma\\[4mm]
\ds\int_{s_1}^{s}\,T_2(\sigma)\delta\theta(\sigma)\,\text{d}\sigma\\[4mm]
\delta\theta(s)
\end{array}
\right].
\end{equation}
The matrix $\mathscr{M}(s)$  is non-singular, and therefore invertible, whenever
\begin{equation}
\label{conjpoints}
\det\left[
\mathscr{M}(s)\right]\neq 0,\qquad \forall\, s\in(s_1,s_2],
\end{equation}
a condition expressing \emph{the  absence of conjugate points within the integration interval}. The latter condition, together with the  \lq Legendre's condition' ($H_1(s)>0$ $\forall \, s$), provides the necessary and sufficient condition for the stability of  fixed length systems \cite{bolza}, namely, when  both the rod's ends are constrained by clamps ($\delta \ell=0$). However,  condition (\ref{conjpoints}) only represents a necessary condition for variable-length systems ($\delta \ell\neq0$), since  perturbations in the length have to be considered.

By considering the isoperimetric constraints (\ref{BCwvarend_iso}), the relations (\ref{newint}) and (\ref{pdiadepprimo}), and the constraint (\ref{eq2}), 
the second variation (\ref{var2bolza0}) reduces to
\begin{equation}
\label{2var_varends}
\delta^2  \mathcal{V}=\int_{s_1}^{s_2}H_1\left(\textbf{p}'\cdot\textbf{u}\right)^2 \text{d}\sigma+\left.\left[\textbf{p}\otimes\textbf{p}:\int_{s_1}^{s}\Psi(\textbf{u})\otimes \textbf{u}\,\text{d}\sigma+\delta\theta\,H_1 (\textbf{p}\cdot \textbf{u}')\right]\right|^{s_2}_{s_1}+\mathscr{F}(s_2)\delta  \ell^2,
\end{equation}
where  the second term on the right hand side further simplifies  to
\begin{equation}
\label{secondterm}
\begin{split}
\left.\left[\textbf{p}\otimes\textbf{p}:\int_{s_1}^{s}\Psi(\textbf{u})\otimes \textbf{u}\,\text{d}\sigma+\delta\theta\,H_1 (\textbf{p}\cdot \textbf{u}')\right]\right|_{s_1}^{s_2}=& q_1(s_2)\int_{s_1}^{s_2}\, T_1\delta\theta \text{d}\sigma+q_2(s_2)\int_{s_1}^{s_2}\, T_2\delta\theta \text{d}\sigma\\[2mm]
&
+\delta\theta(s_2)H_1(s_2)\,\textbf{p}(s_2)\cdot \textbf{u}'(s_2),
\end{split}
\end{equation}
where  $\textbf{u}'(s_2)$ is known from solving (\ref{functs}), while  $\textbf{p}(s_2)$ can be obtained by particularizing the linear system 
(\ref{systemm}) at the coordinate $s_2$,
\begin{equation}
\label{systemms2}
\mathscr{M}(s_2)
\textbf{p}(s_2)=\textbf{z}(s_2).
\end{equation}
Noting that, by applying the compatibility condition (\ref{BCwvarend_bc})$_2$ and the two isoperimetric constraints (\ref{BCwvarend_iso}), $\textbf{z}(s_2)$ is given by
\begin{equation}
\textbf{z}(s_2)=\tilde{\textbf{z}}(s_2)\delta \ell, \qquad \mbox{where}\qquad \tilde{\textbf{z}}(s_2)=\left[
\begin{array}{ccc}
f_1(s_2)\\
f_2(s_2)\\
W(s_2)
\end{array}
\right],
\end{equation}
and, defining the unknown vector $\textbf{p}(s_2)$ as 
\begin{equation}
\textbf{p}(s_2)=\tilde{\textbf{p}}(s_2)\delta  \ell,
\end{equation}
the  linear system (\ref{systemms2}) can be rewritten as
\begin{equation}
\label{systemms2b}
\mathscr{M}(s_2)
\tilde{\textbf{p}}(s_2)=\tilde{\textbf{z}}(s_2),
\end{equation}
so that the second variation $\delta^2  \mathcal{V}$ simplifies as the following quadratic form, 
\begin{equation}
\label{2var_varends2}
\delta^2  \mathcal{V}=\int_{s_1}^{s_2}H_1\left(\textbf{p}'\cdot\textbf{u}\right)^2 \text{d}\sigma-\Xi \, \delta  \ell^2,
\end{equation}
where
\begin{equation}
    \Xi=-\Bigl\{\mathscr{F}(s_2)+\tilde{\textbf{p}}(s_2)\cdot [0,f_1(s_2),f_2(s_2)]^\intercal+W(s_2) H_1(s_2)\,\tilde{\textbf{p}}(s_2)\cdot \textbf{u}'(s_2)\Bigr\}.
\end{equation} 

The positive definiteness of the second variation $\delta^2 \mathcal{V}$, Eq. (\ref{2var_varends2}), for every compatible perturbations $\delta\theta(s)$ and $\delta\ell$ provides the necessary and sufficient conditions for the stability of an equilibrium configuration for a system with one variable endpoint subject to two isoperimetric constraints. When the Legendre's condition ($H_1(s)>0$) holds, the stability criterion for variable-length systems is  represented by
\begin{equation}
\label{necsuffconds}
\Xi< 0,\qquad \mbox{and}\qquad\det\left[
\mathscr{M}(s)\right]\neq 0,\qquad \forall\, s\in(s_1,s_2].
\end{equation}

\subsection{Stability criterion for the \lq elastica sling'}
With  reference to the \lq \emph{elastica sling}' system under consideration, its second variation (\ref{secondvarsling}), the compatibility condition (\ref{compatibility})$_2$ (with $k=1$), and the first order expansion of the isoperimetric constraints (\ref{isoperimetricderived}) can be obtained from the generic expressions   (\ref{var2bolza0}), (\ref{BCwvarend_bc})$_2$, and (\ref{BCwvarend_iso}) 
 through the following relations 
\begin{equation}
\label{changes}
\begin{array}{lllll}
H_1(s)=B>0,
& T_1(s)=-\sin\theta(s),
& f_1(s_2)=-\cos\theta_2,
& W(s_2)=-\theta'(s_2),
\\[2mm]
H_2(s)=-R_x \cos\theta(s)-R_y \sin\theta(s),
& T_2(s)=\cos\theta(s),
& f_2(s_2)=-\sin\theta_2,
& \mathscr{F}(s_2)=-B\theta'(s_2)\theta''(s_2),
\end{array}
\end{equation}
which imply
\begin{equation}
\tilde{\textbf{z}}(s_2)=-\left[
\begin{array}{ccc}
\cos\theta_2\\
\sin\theta_2\\
\theta'(s_2)
\end{array}
\right].
\end{equation}
The stability for the \lq \emph{elastica sling}' can be  assessed through the stability condition (\ref{necsuffconds})  by considering the corresponding square matrix $\mathscr{M}(s)$
\begin{equation}
\label{matrix22}
\mathscr{M}(s)=\left[
\begin{array}{ccc}
\ds-\int_{s_1}^{s} u(\sigma)\sin\theta(\sigma)\,\text{d}\sigma & \ds-\int_{s_1}^{s} v_1(\sigma)\sin\theta(\sigma)\,\text{d}\sigma &\ds -\int_{s_1}^{s} v_2(\sigma)\sin\theta(\sigma)\,\text{d}\sigma\\[5mm]
\ds\int_{s_1}^{s} u(\sigma)\cos\theta(\sigma)\,\text{d}\sigma & \ds\int_{s_1}^{s} v_1(\sigma)\cos\theta(\sigma)\,\text{d}\sigma & \ds\int_{s_1}^{s} v_2(\sigma)\cos\theta(\sigma)\,\text{d}\sigma\\[5mm]
u(s) & v_1(s) & v_2(s)
\end{array}
\right],
\end{equation}
and  parameter $\Xi$
\begin{equation}
\Xi=B\theta'(s_2)\theta''(s_2)+\tilde{\textbf{p}}(s_2)\cdot [0,\cos\theta_2,\sin\theta_2]^\intercal+B\theta'(s_2)\,\tilde{\textbf{p}}(s_2)\cdot \textbf{u}'(s_2).
\end{equation}
\paragraph{Critical case $\Xi=0$.} It is finally observed that  $\Xi=0$ represents a critical condition because   the stability criterion (\ref{necsuffconds}) is no longer met. Although a rigorous analytical assessment of the  stability is not addressed here for the critical equilibrium configurations, through consideration of an auxiliary clamped-clamped system, all the several performed semi-analytical analyses detected an unstable character. More specifically, as  shown in the next Section,  it is  found that whenever $\Xi=0$  then
\begin{equation}
\label{variamicc}
    \left.\delta^2\mathcal{V}\right|_{\delta \theta=\widehat{\delta \theta}}=0
    \quad
    \mbox{and}
    \quad
   \left. \delta^3\mathcal{V}\right|_{\delta \theta=\widehat{\delta \theta}}\neq0
    \qquad
    \mbox{for}
    \quad
   \widehat{\delta \theta}(s)=\left.\dfrac{\partial\theta^{cc}_{eq}(s,\ell)}{\partial \ell} \right|_{\ell=\ell_{eq}}\delta\ell,
\end{equation}
where $\theta^{cc}_{eq}(s,\ell)$ is the equilibrium rotation field for the clamped-clamped system with rod's length $\ell$ subject to the same boundary conditions (\ref{constraints0}) and (\ref{constraints1}) of  the variable-length system. Therefore, based on this observation,
\begin{center}
\emph{
the non-trivial equilibrium configurations for the variable-length system\\ providing the critical condition $\Xi=0$ are  found unstable.}
\end{center}

\section{Results}
\subsection{Theoretical predictions}

Once the planar equilibrium configurations are obtained, their stability can be assessed through the criterion provided by Eq. (\ref{necsuffconds}). A main result of the present procedure  is that no more than one stable equilibrium configuration exists for the variable-length system. Moreover,  all the  equilibrium configurations with more than one inflection point ($m>1$) are found unstable in the presence of a sliding sleeve, differently 
from the double-clamped system where  all the configurations with more than two inflection points ($m>2$) are unstable \cite{love}.

\begin{figure}[!h]
\renewcommand{\figurename}{\footnotesize{Fig.}}
    \begin{center}
\includegraphics[width=0.95\textwidth]{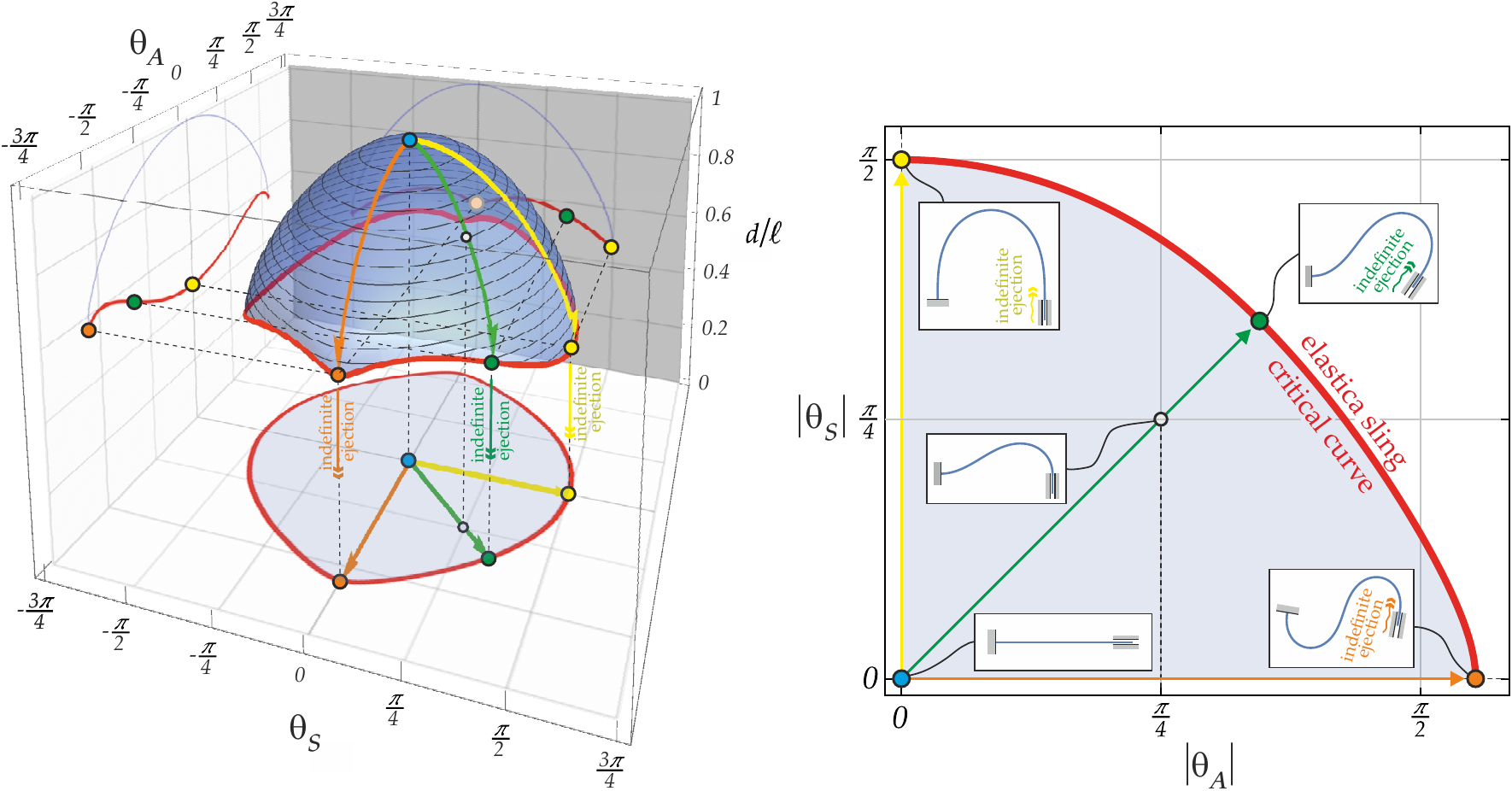}
\caption{\footnotesize{(Left) Surface representing the dimensionless value $d/\ell$ of the stable equilibrium configuration with varying  the end rotations $\theta_A$ and $\theta_S$  (dataset available as Supplementary material). The red boundary represents the critical states leading to indefinite ejection of the elastic rod and which are characterized by a vertical tangent of the equilibrium surface. (Right) Due to symmetry properties, a quarter of the proposed critical curve is represented (red) together with some deformed shapes at equilibrium. Attaining the border leads to the indefinite ejection of the elastic rod from the constraints.}
		}
    \label{fig_eject}
    \end{center}
\end{figure}

The theoretical surface collecting the  normalized distance $d/\ell$ of the unique stable equilibrium configuration is reported in Fig. \ref{fig_eject} (left) in terms of $\theta_A$ and $\theta_S$, respectively defined as antisymmetric and symmetric parts of the  inclinations $\theta_1$ and $\theta_2$ imposed at the two sliding sleeves,
\beq
\theta_A=\dfrac{\theta_1+\theta_2}{2},\qquad
\theta_S=\dfrac{\theta_2-\theta_1}{2},
\eeq
where, considering the necessary condition (\ref{bound}), the modulus of the symmetric angle $\theta_S$ for the existence of an equilibrium configuration is limited by
\beq\label{boundthetas}
\left|\theta_S\right|\leq\dfrac{\pi}{2}.
\eeq
The equilibrium surface of Fig. \ref{fig_eject} (left) is obtained through a
dataset of  66,400  triads $\{d/\ell,\theta_A,\theta_S\}$ (available as Supplementary material) evaluated through a
semi-analytical procedure based on the numerical solution $d/\ell$ of the nonlinear system of equations (\ref{solselastica}) by varying $\theta_A$ and $\theta_S$, for which the stability criteria (\ref{necsuffconds}) holds.

The stable equilibrium surface is bounded by a closed (red) curve defining the critical triads $\{d/\ell, \theta_A, \theta_S\}$ for which the rod is indefinitely ejected because the modulus of the gradient of $d/\ell$ approaches an infinite value and an  unbounded growth of the rod's length $\ell$, therefore realizing an \lq \emph{elastica sling}'. 
With reference to the mathematical conditions of stability, the \lq \emph{elastica sling}' critical curve is associated with solutions for which the inequality condition (\ref{necsuffconds})$_1$ is no longer met because $\Xi=0$, while the determinant condition (\ref{necsuffconds})$_2$  still holds. No equilibrium configuration is  found for rotation pairs  outside of the critical  curve, being such rotations incompatible  with the \lq mother curve'  for the \lq \emph{elastica sling}' (Sect. \ref{eqconfs}, Fig. \ref{fig_mothercurve}).  In practice,  the critical curve has been defined through a  bisection algorithm by considering intervals with bounds provided by one pair of $\theta_A$ and $\theta_S$ for which the stable solution exists and by another pair for which no solution exists. Convergence to the critical condition is considered to be reached when  the difference in both angles combination is smaller than $10^{-6}$.

The projection of the \lq \emph{elastica sling}' critical curve in the $\theta_A$--$\theta_S$ rotation plane is reported in   Fig. \ref{fig_eject} (right). Due to its symmetry, the projection of the critical curve is reported only for one quarter,  representative of the absolute values $|\theta_A|$--$|\theta_S|$ of the critical pairs. Three specific rotation paths (at constant ratio $\theta_S/\theta_A$) are  highlighted and complemented by the equilibrium configuration at some specific rotation pairs. More specifically, the highlighted paths are  for $\theta_A=0$ (yellow), for $\theta_S=0$ (orange), and for $\theta_A=\theta_S$ (green), showing that only the first path involves deformed configurations with no inflection points ($m=0$), while one inflection point ($m=1$) is present in the other two paths.

\begin{figure}[!h]
\renewcommand{\figurename}{\footnotesize{Fig.}}
    \begin{center}
   \includegraphics[width=1\textwidth]{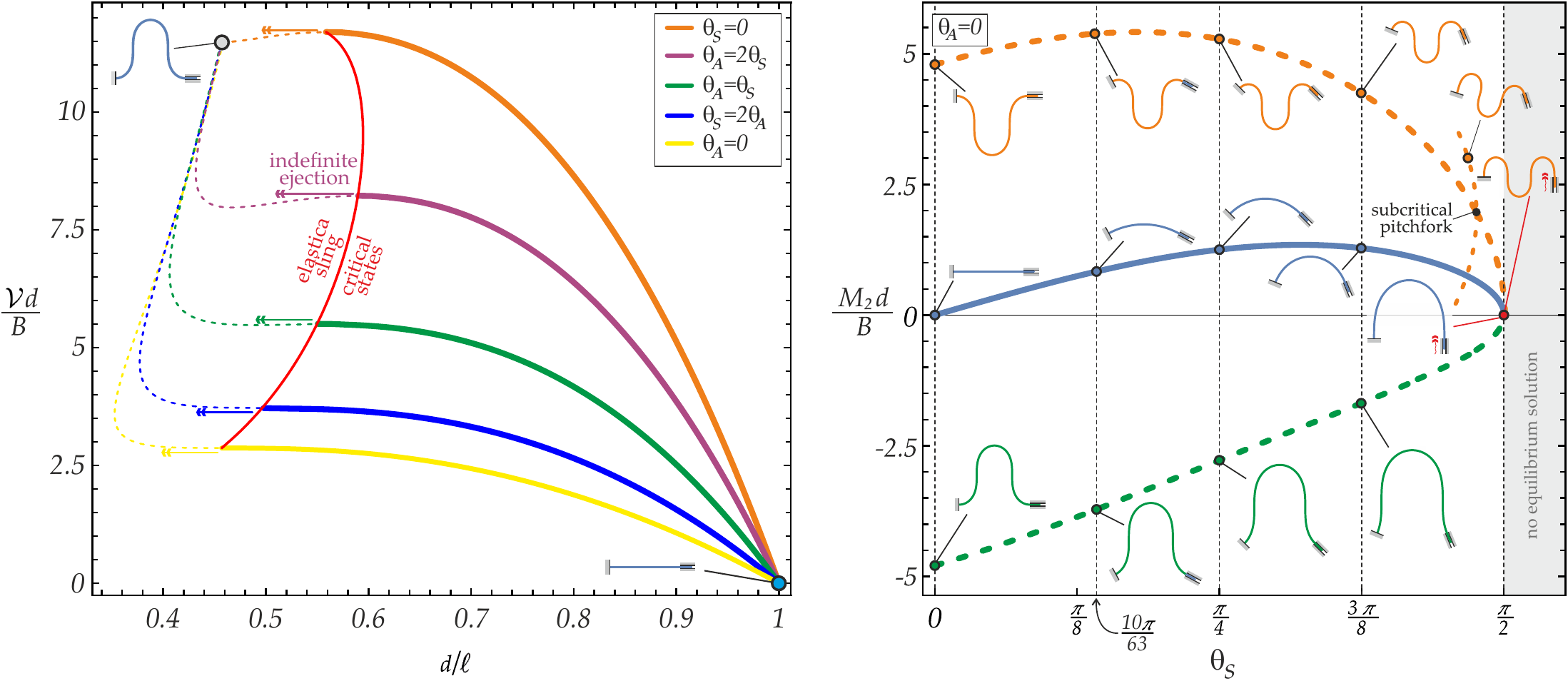}
   \caption{\footnotesize{(Left) Total potential energy $\mathcal{V}$ for an elastica constrained through a clamp and a sliding sleeve as function of $d/\ell$. Stable (continuous lines) and unstable  (dashed lines) paths are reported for different ends' rotation path, in particular the cases $\theta_A=0$, $\theta_A=\theta_S$ and $\theta_S=0$ are reported in yellow, green and orange, respectively, and represent the three equilibrium paths reported in Fig. \ref{fig_eject}. Indefinite ejection occurs when the system attains a local maximum in the total potential energy $\mathcal{V}$.  (Right) Stable (continuous line) and unstable (dashed line) evolutions of the  bending moment $M_2=B \theta'(s_2)$ (made dimensionless through division by $B/d$) with increasing symmetric rotation $\theta_S$ at  null antisymmetric angle $\theta_A=0$. Related equilibrium configurations are highlighted. In the particular case $\theta_S \simeq 10 \pi/63$, the related deformed shapes are those represented in Fig. \ref{fig_stab} (right, above). }}
    \label{fig_tpe}
    \end{center}
\end{figure}

The total potential energy $\mathcal{V}$ for five different straight paths in the  $\theta_A$--$\theta_S$ plane  is reported in Fig. \ref{fig_tpe} (left) for decreasing  value of the ratio $d/\ell$. Stable (unstable) configurations are reported as continuous (dashed) lines. All of these five paths initiate at null rotations and with a straight configuration ($d/\ell=1$) and lose stability at the corresponding critical value $d/\ell$ of ejection,  where the total potential energy $\mathcal{V}$ attains a local maximum. It is also observed that, among the different paths, the highest (lowest) total potential energy $\mathcal{V}$ at  the critical  value of $d/\ell$ is stored in the system evolving with a pure antisymmetric (symmetric) rotation path, $\theta_S=0$ ($\theta_A=0$), therefore corresponding to the most (less) propulsive case of indefinite ejection. The \lq \emph{elastica sling}' critical states (red continuous line,  projection of the \lq \emph{elastica sling}' critical curve reported in Fig. \ref{fig_eject}, left) collects the indefinite ejection conditions corresponding to all the possible continuous evolution in the rotations of the two sliding sleeves starting from the straight undeformed configuration of the rod. It is further noted that all the different unstable equilibrium paths (dashed lines) have a common state (grey point), corresponding to the symmetric configuration for $\theta_A=\theta_S=0$ with two inflection points ($m=2$). These  unstable paths can be attained by further decreasing the rotations' amplitude after reaching the critical configuration and moving along a different equilibrium path.

The bending moment at the second sliding sleeve, $M_2=B\theta'(s_2)$, versus the imposed symmetric  rotation $\theta_S$  is reported in Fig. \ref{fig_tpe} (right) for  null antisymmetric rotation ($\theta_S=\theta_2=-\theta_1$, $\theta_A=0$). The response is reported for the unique stable path configuration as continuous line and for  two unstable paths  as dashed lines. 
The three equilibrium paths join together at the point $\theta_S=\pi/2$, where a vertical tangent is displayed for each one of these and indefinite ejection takes place. No (stable or unstable) equilibrium configuration exists for $\theta_S>\pi/2$ (grey region) since  condition (\ref{boundthetas}) is no longer met.
Deformed configurations at specific values of the angle $\theta_S$ are included along the moment-rotation evolution, showing that the stable configurations have no inflection points ($m=0$) while all the unstable ones have  two inflection points ($m=2$). Interestingly, the stable path associated with the non-inflectional elastica loses the stability at $\theta_S=\pi/2$ as two inflection points simultaneously arise at the two edge coordinates, $s_1$ and $s_2$.  

\begin{figure}[!htb]
\renewcommand{\figurename}{\footnotesize{Fig.}}
    \begin{center}
   \includegraphics[width=0.95\textwidth]{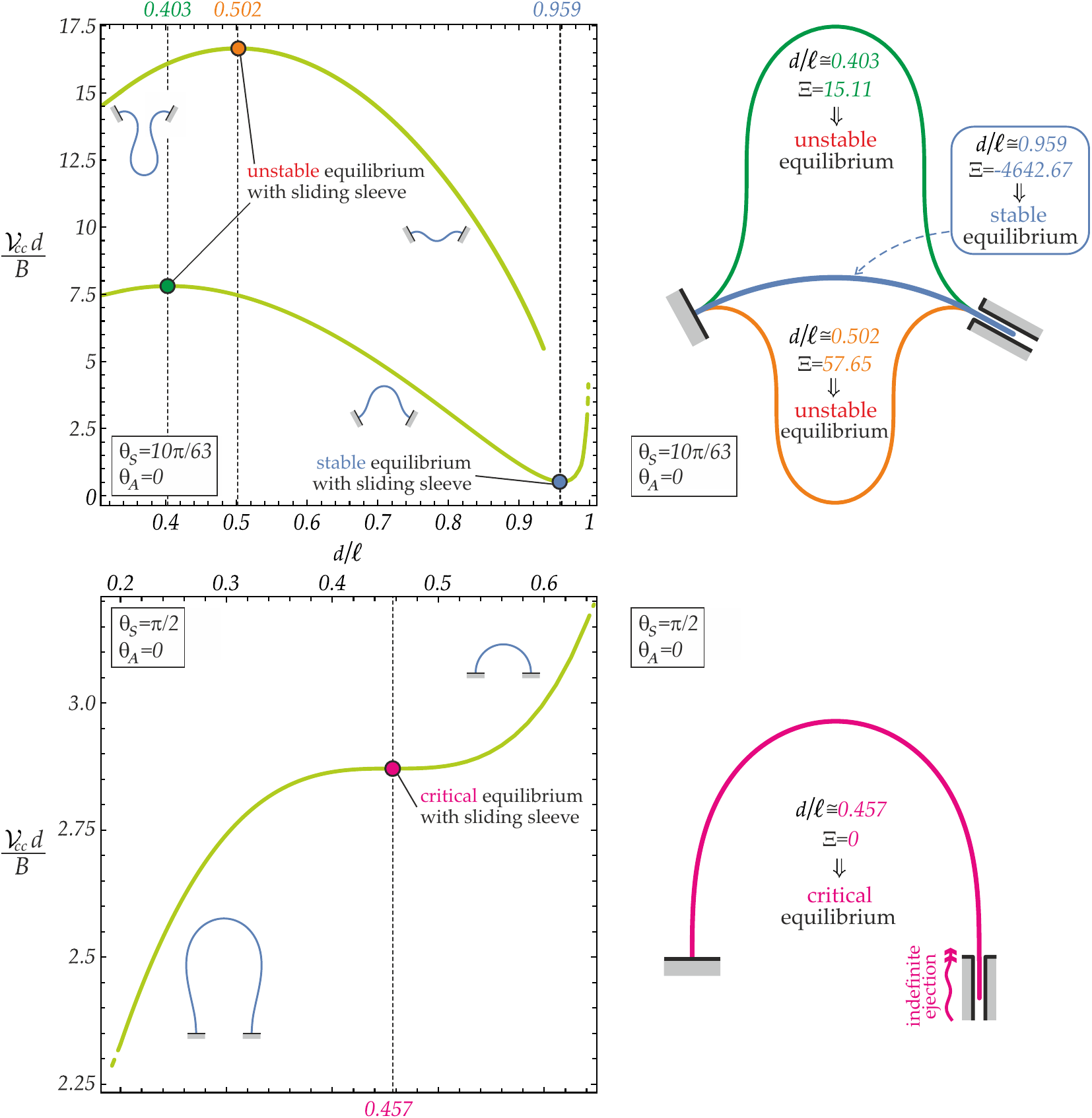}
    \caption{\footnotesize{(Left) Total potential energy $\mathcal{V}_{cc}$ of the double-clamped system of fixed rod's length $\ell$ at equilibrium  with varying the ratio $d/\ell$. The two clamp rotations are considered such that $\theta_A=0$ and $\theta_S=10\pi/63$ (top left) or $\theta_S=\pi/2$ (bottom left).
    Stationary points of $\mathcal{V}_{cc}$ define the equilibrium solution for the system where at least one of the two clamps is replaced by a sliding sleeve.
 (Right) Corresponding deformed shapes at (stable, unstable, and critical)  equilibrium  for the variable-length system, corresponding to the stationary points on the $\mathcal{V}_{cc}$ curve (highlighted through a circle with the same color of the deformed shape).}
		}
    \label{fig_stab}
    \end{center}
\end{figure}

To further understand the present results and in particular how an equilibrium configuration  is generated,  when this is stable or unstable, and when this represents a critical condition of ejection, an analogous system with imposed rotations $\theta_A$ and $\theta_S$ but with clamps at both ends is considered. The total potential energy $\mathcal{V}_{cc}$ of the clamped-clamped system at equilibrium with a rod's  length $\ell$  is reported in Fig. \ref{fig_stab} (left)  with varying the ratio $d/\ell$ for $\theta_A=0$ and $\theta_S=10\pi/63$ (above, left) and $\theta_S=\pi/2$ (bottom left).
It is interesting to note that 
equilibrium configurations for the system with a sliding sleeve correspond to those ratios $d/\ell$ providing the stationary condition of $\mathcal{V}_{cc}$ for the clamped-clamped system, $\partial\mathcal{V}_{cc}/\partial{(d/\ell)}=0$. More specifically, the stable (unstable) configuration of the sliding sleeve system corresponds to the local minimum (maximum) of the total potential energy $\mathcal{V}_{cc}$ of the clamped-clamped system (Fig. \ref{fig_stab}, above left). Moreover, the critical condition of indefinite ejection for the system with a sliding sleeve corresponds to a saddle point of $\mathcal{V}_{cc}$ (Fig. \ref{fig_stab}, bottom left). Similarly to all the equilibrium configurations belonging to the  \lq \emph{elastica sling}' critical curve, the considered critical  configuration is found  to provide $\Xi=0$ and  assessed as unstable because, while the  second derivative is null ($\partial^2\mathcal{V}_{cc}/\partial{(d/\ell)^2}=0$), the  third derivative does not vanish ($\partial^3\mathcal{V}_{cc}/\partial{(d/\ell)^3}\neq0$),  providing  the respective variations $\delta^2\mathcal{V}=0$ and $\delta^3\mathcal{V}\neq0$ under the specific rotation perturbation $\delta\theta(s)$  as described by Eq. (\ref{variamicc}).

\subsection{Experimental validation}

\begin{figure}[!h]
\renewcommand{\figurename}{\footnotesize{Fig.}}
    \begin{center}
   \includegraphics[width=0.98\textwidth]{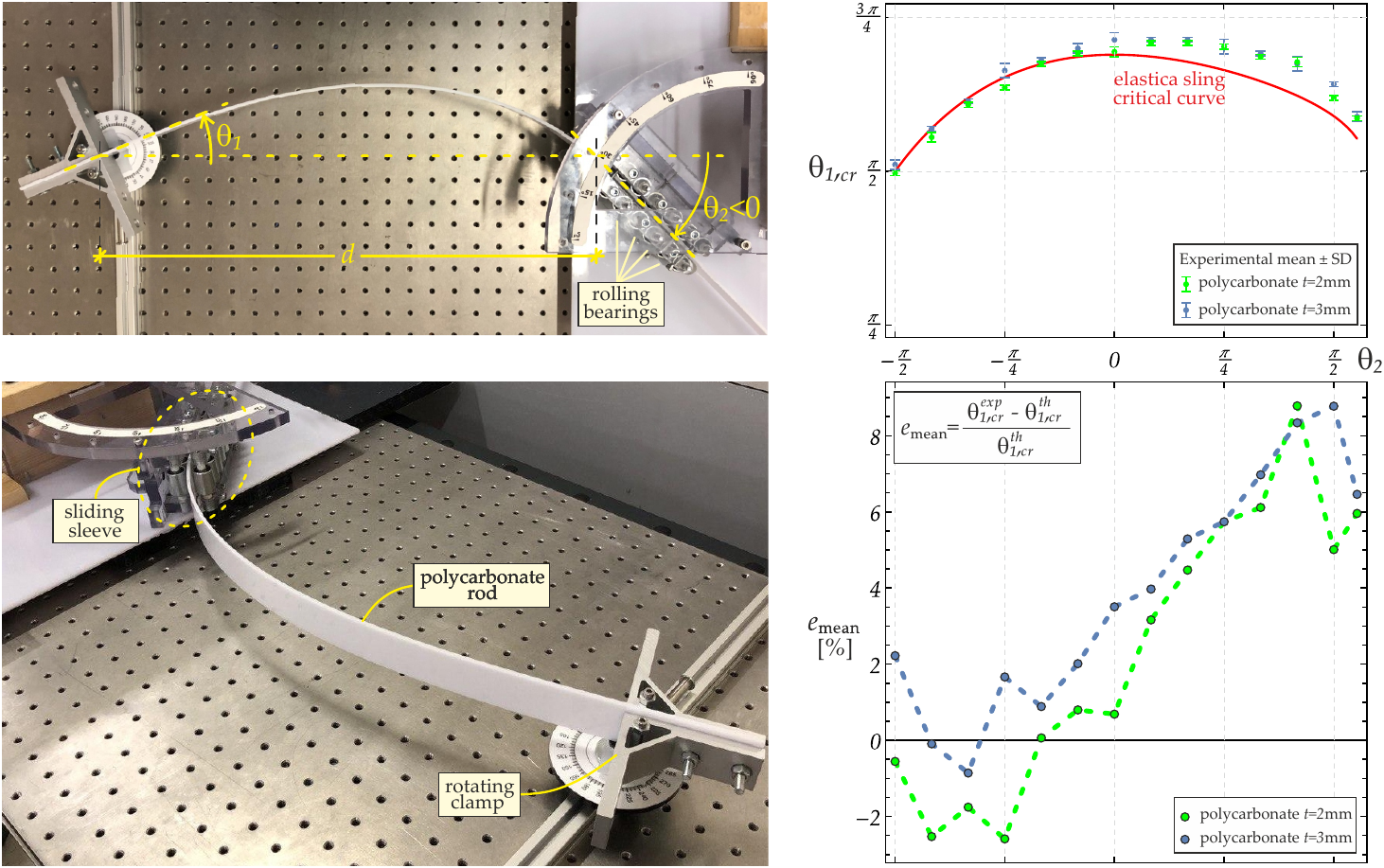}
    \caption{\footnotesize{(Left) Experimental setup realized through a polycarbonate rod (white) clamped at one end and constrained at the other  with a  sliding sleeve.  (Right, above) Experimental measures of the critical angle $\theta_{1,cr}^{exp}$, obtained by testing two polycarbonate rods with different thickness $t$,
    compared with the theoretical prediction $\theta_{1,cr}^{th}$ from the \lq \emph{elastica sling}' critical curve (red curve, projection in the $\theta_{1}-\theta_{2}$ plane of the critical curve in Fig. \ref{fig_eject}). (Right, bottom) Corresponding percentage error $e_{\text{mean}}$ between the experimental measures of the critical angle $\theta_{1,cr}^{exp}$ and the corresponding theoretical value $\theta_{1,cr}^{th}$.}
		}
    \label{fig_exps}
    \end{center}
\end{figure}

The obtained theoretical predictions are finally   validated through experiments on  polycarbonate rods (Young's Modulus $\mathrm{E}$ = 2350 MPa and volumetric mass density $\rho$ = 1180 kg/m$^3$) constrained by a rotating clamp at $s_1$ and a sliding sleeve at $s_2$, Fig. \ref{fig_exps} (left).
The sliding sleeve exploited here is the same adopted in \cite{bosiarmscale} for realizing the elastica arm scale. The tests are conducted with fixed values of  $\theta_2$ and  increasing values of $\theta_1$, controlled through a slow rotation  by hand of  the clamp. When the rod is observed to have an uncontrolled ejection, the rotation $\theta_1$ is stopped and the measured value is recorded as $\theta_{1,cr}^{exp}(\theta_2)$. The comparison between the experimental measure of the critical angle $\theta_{1,cr}^{exp}$ versus its theoretical prediction $\theta_{1,cr}^{th}$ is  reported  as a function of the angle $\theta_2$ in Fig.  \ref{fig_exps} (right, above). The corresponding  error $e_{\text{mean}}=\left(\theta_{1,cr}^{exp}-\theta_{1,cr}^{th}\right)/\theta_{1,cr}^{th}$ is also shown  in Fig. \ref{fig_exps} (right, bottom).
Two different thicknesses for the polycarbonate rod of width  20mm are considered, $t=2$mm  and $t=3$mm.  The distance between the two constraints is fixed to $d=$500mm. It is observed that the error is greater for $\theta_2>0$ than that for $\theta_2<0$ because corresponding to deformed shapes with  larger rotations. When   $\theta_2>0$ the maximum error $e_{\text{mean}}\approx9\%$   is achieved for  $\theta_2\in[5/6,1]\pi/2$, while  when   $\theta_2<0$ the maximum error $e_{\text{mean}}\approx4\%$   is achieved for  $\theta_2\approx 0$. Therefore, it follows that the error $e_{\text{mean}}$ increases as the ratio $\theta_A/\theta_S$ increases, to which corresponds an increase of the total potential energy $\mathcal{V}$ at the critical conditions (cfr. Fig. \ref{fig_tpe}). This observation is in agreement with a previous experimental experience in a different setting \cite{cazzolliECM} where polikristal rods, displaying similar properties of the polycarbonate ones, were tested. Nevertheless, apart from such discrepancies which could be reduced by adopting a rod's material characterized by smaller intrinsic viscosity and weight-stiffness ratio, the theoretical curve is very well capturing the trend of  the  experimental measures.

\begin{figure}[!h]
\renewcommand{\figurename}{\footnotesize{Fig.}}
    \begin{center}
   \includegraphics[width=.85\textwidth]{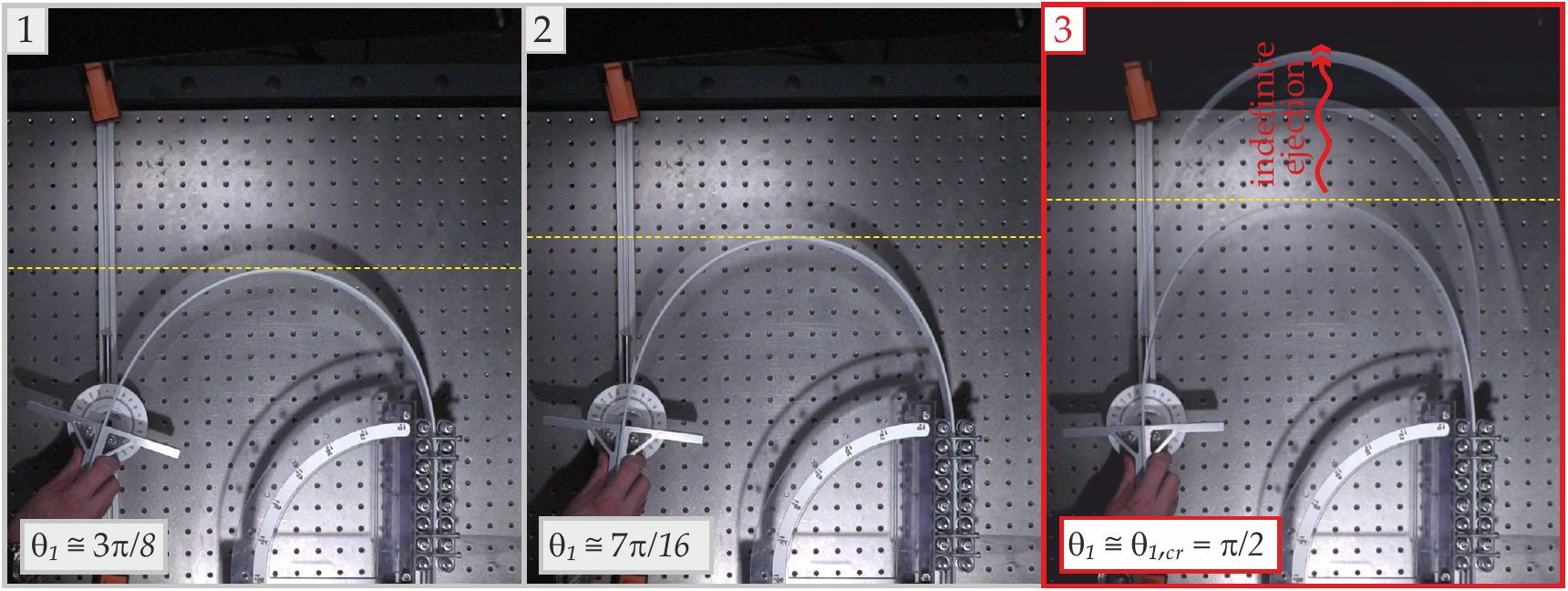}
    \caption{\footnotesize{Snapshots taken during the experiment  under $\theta_2=-\pi/2$  at increasing $\theta_1$ (from left to right). 
    The  indefinite ejection of the rod from the sliding sleeve is displayed for  $\theta_1\approx\theta_{1,cr}=\pi/2$, according to the theoretical predictions.}
		}
    \label{fig_frames}
    \end{center}
\end{figure}

Furthermore, the evolution of the system  with $\theta_2=-\pi/2$ is reported in Fig. \ref{fig_frames} through three snapshots taken at increasing $\theta_1$. According to the theoretical predictions, a fast   motion can be appreciated through the indefinite rod's ejection  when $\theta_1$ approaches the critical value, $\theta_{1,cr}(\theta_2=-\pi/2)=\pi/2$.

\section{Conclusions}
The equilibrium and the stability have been addressed for a planar mechanical system, based on a  elastic rod constrained at its two edges by a pair of sliding sleeves with  controlled position and inclination.
The presence of the sliding sleeve  constraints defines a variable-length   system subject to configurational forces at its edges. The nonlinear equations governing the equilibrium are derived through a variational approach and  solved  in terms of elliptic integrals. \\ 
For the first time, a theoretical stability criterion has been defined  for  variable domain systems, by extending a previous framework restricted to  fixed domains. The stability analysis shows that no more than one stable equilibrium configuration  exists  for every sliding sleeves’ inclination pairs, that is characterized by none or one inflection point. The set of critical conditions giving rise to the  loss of stability is revealed, for which  the elastic element is indefinitely ejected from the constraints, thus realizing   an \lq \emph{elastica sling}'. Finally, the theoretical results are  validated by  comparison with experimental tests carried out on a physical prototype.

As the elastic energy stored in the variable-length element is released and converted into kinetic energy when 
the mechanical instability occurs, the present system can be exploited as a novel actuation mechanism in several technological fields. Examples of application include soft robotic locomotion, wave mitigation and energy harvesting.

\paragraph{CRediT authorship contribution statement.}
A. Cazzolli: Formal analysis; Investigation; Methodology; Software; Data curation; Validation; Visualization; Writing -- original draft; Writing -- review and editing. 
F. Dal Corso: Conceptualization; Project administration; Formal analysis; Investigation; Methodology; Supervision; Visualization; Writing -- original draft, Writing -- review and editing; Funding acquisition.

\paragraph{Acknowledgements.}
The authors gratefully  acknowledge financial support from the ERC advanced grant
ERC-ADG-2021-101052956-BEYOND.
The authors also acknowledge the Italian Ministry of Education, Universities and Research (MUR) in the framework of the project DICAM-EXC (Departments of Excellence 2023-2027, grant L232/2016).
This work has been developed under the auspices of INDAM-GNFM.

\appendix
\renewcommand{\theequation}{{\thesection.}\arabic{equation}}
\renewcommand\thefigure{\thesection.\arabic{figure}}   

\section{General variational approach for variable domains}\label{AppendixA}
\setcounter{equation}{0}
\setcounter{figure}{0} 

\subsection{Preliminaries}
The generic smooth variation $s^*$ for the curvilinear coordinate $s$ and $\theta^*$ for the rotation field $\theta$ are introduced for describing the perturbed solution of the rod outside the sliding sleeves 
\begin{equation}
s^*=\mathcal{S}(s,\theta,\theta';\epsilon),\quad \theta^*=\mathcal{T}(s,\theta,\theta';\epsilon),
\end{equation}
with $\mathcal{S}(s,\theta,\theta';0)=s$ and $\mathcal{T}(s,\theta,\theta';0)=\theta$ for $\epsilon=0$. A first-order  expansion of the previous equations for small values of $\epsilon$  provides
\begin{equation}
\label{linexps}
s^*= s+\epsilon \phi(s,\theta,\theta')+o(\epsilon),\quad \theta^*= \theta(s)+\epsilon \psi(s,\theta,\theta')+o(\epsilon),
\end{equation}
where $\phi$ and $\psi$ are the first derivatives of the functions $\mathcal{S}$ and $\mathcal{T}$  with respect to $\epsilon$ and evaluated at $\epsilon=0$. The perturbations follow as
\begin{equation}
\label{deltafunz}
\widetilde{\Delta s}=s^*-s=\epsilon \phi(s) +o(\epsilon),\qquad \widetilde{\Delta \theta}= \theta^*(s^*)-\theta(s)=\epsilon \psi(s) +o(\epsilon),
\end{equation}
from which, following Gelfand and Fomin\cite{gelfand}, the  first-order variations $\widetilde{\delta \theta}$ and $\widetilde{\delta s}$ are given
\begin{equation}
\label{truevars}
\widetilde{\delta s}=\epsilon \phi,\qquad \widetilde{\delta\theta}=\epsilon \psi.
\end{equation}
The notation \lq $\widetilde{\, \cdot \,}$' is introduced to distinguish the \lq general' variations proposed by Gelfand and Fomin \cite{gelfand} from those proposed in Eqs. (\ref{perturb0}) - (\ref{legameFDC}). It is also worth to underline that, in the case of Dirichlet boundary conditions given in terms of the field $\theta(s)$ at both ends, the fixed ends' rotations must be fulfilled for both the perturbed and unperturbed configurations, so that the condition $\theta ^*(s_i^*)=\theta(s_i)$ holds at each endpoint $s_i$ ($i=1,2$) and therefore
\beq
\label{dirichconds}
\widetilde{\delta\theta}(s_i)=0 \qquad i=1,2,
\eeq
while the perturbation $\widetilde{\delta s}$ is in general different from zero.

A further rotation difference $\overline{\Delta\theta}$ introduced by Elsgolts \cite{elsgolts} and reported in Eq. (\ref{perturba2}) can be rewritten as: 
\begin{equation}
\label{pippo}
\overline{\Delta\theta}= \theta^*(s)-\theta(s)=\epsilon \overline\psi(s) +o(\epsilon),
\end{equation}
which represents the gap between the rotation of the perturbed solution and of the equilibrium evaluated  at a fixed coordinate $s$, so that its related variation is 
\begin{equation}
\overline{\delta\theta}=\epsilon \overline\psi.
\end{equation}
The \lq general' difference introduced in Eq. (\ref{deltafunz})$_2$ can be therefore rewritten as
\begin{equation}
\widetilde{\Delta \theta}= \theta^*(s^*)-\theta^*(s)+\overline{\Delta\theta},
\end{equation}
and  expanded in Taylor series for a small difference $\Delta s$ as
\begin{equation}
\epsilon \psi +o(\epsilon)\simeq \sum_{i=1}^{\infty} \left[\frac{\partial^i\theta^*(s)}{\partial s^i }\frac{(s^*-s)^i}{i!}\right]
+\epsilon\overline{\psi}+o(\epsilon).
\end{equation}
By truncating the expansions in the expressions of $\widetilde{\Delta \theta}$,  $\overline{\Delta\theta}$ and $\widetilde{\Delta s}$  to the first order (so that $\theta^*(s)=\theta(s)+\epsilon \overline\psi$) the following expressions are obtained for increasing orders
\begin{equation}
\label{expansions}
\widetilde{\delta\theta}=\theta'\widetilde{\delta s} +\overline{\delta\theta},\quad 0=\frac{\theta''}{2}\widetilde{\delta s}^2 +\overline{\delta\theta}'\widetilde{\delta s},\quad 0=\frac{\theta'''}{6}\widetilde{\delta s}^3 +\frac{\overline{\delta\theta}''}{2}\widetilde{\delta s}^2,\quad\dots
\end{equation}
By means of Eqs. (\ref{dirichconds}) enforcing Dirichlet boundary conditions at both ends and Eq. (\ref{expansions})$_1$, one can express the values assumed by $\overline{\delta\theta}$ at both ends in the following form
\beq
\label{dirichcondsthbarra}
\overline{\delta\theta}(s_i)=-\theta'(s_i)\widetilde{\delta s}_i, \qquad \forall \, s_i.
\eeq

Moreover,  considering the following difference of derivatives
\begin{equation}
\label{truedervars1}
\widetilde{\Delta \theta_s}= \frac{\partial \theta^*(s^*)}{\partial s^*}-\frac{\partial \theta(s)}{\partial s},
\end{equation}
the following linear equation is obtained from its first order expansion \cite{gelfand}
\begin{equation}
\label{truedervars}
\widetilde{\delta\theta_s}=\theta''\widetilde{\delta s} +\overline{\delta\theta}',
\end{equation}
which is therefore different from $\widetilde{\delta\theta}'$, the simple derivative of the perturbation $\widetilde{\delta\theta}$,
\begin{equation}
\label{truedervars}
\widetilde{\delta\theta_s}\neq \widetilde{\delta\theta}'=\theta''\widetilde{\delta s}+ \theta'\widetilde{\delta s}' +\overline{\delta\theta}'.
\end{equation}

\subsection{General variations of a functional}
\label{generalvars}

For a  functional $\mathcal{V}$ given as the integral of a functional $F$ dependent on the fields $\theta$ and $\theta'$ and subject to Dirichlet boundary conditions at both ends, $\theta(s_1)=\theta_1$ and $\theta(s_2)=\theta_2$, explicitly dependent\footnote{For instance, in the case of an Euler's elastica through an explicitly defined expression of the bending stiffness $B(s)$.} on the curvilinear coordinate $s$ and subject to isoperimetric constraints, whose Lagrange multipliers are collected in the vector $\boldsymbol{\chi}$, 
\begin{equation}
\label{generalfunctional}
\mathcal{V}=\int_{s_1}^{s_2}F\left[s,\theta(s),\theta'(s),\boldsymbol{\chi}\right]\,\text{d}s,
\end{equation}
the first variation $\delta \mathcal{V}$ is
\begin{equation}
\begin{split}
\label{1varcomplete}
\delta \mathcal{V}=&\int_{s_1}^{s_2} \left[\overline{\delta \theta}\left(F_\theta -\frac{\partial}{\partial s}\left(F_{\theta'}\right)\right)+\nabla_{\boldsymbol{\chi}}F\cdot \delta\boldsymbol{\chi}\right]\text{d}s+\left.\left[F\widetilde{\delta s}+F_{\theta'}\overline{\delta \theta}\right]\right|_{s_1}^{s_2}.
\end{split}
\end{equation}
By assuming that the functional $F$ is not explicitly dependent on $s$, the second $\delta^2 \mathcal{V}$ and third $\delta^3 \mathcal{V}$ variations are
\begin{equation}
\begin{split}
\label{23varcomplete}
\delta^2 \mathcal{V}=&\int_{s_1}^{s_2}\left[\overline{\delta\theta}'^2 F_{\theta'\theta'}+\overline{\delta\theta}^2 F_{\theta\theta}+2\,\overline{\delta\theta}\,\overline{\delta\theta}' F_{\theta\theta'}
\right]\text{d}s+\left.\left[ F_s\, -\theta'\,F_\theta\right]\widetilde{\delta s}^2\right|_{s_1}^{s_2}\\
&+2\left[\int_{s_1}^{s_2}\left[\nabla_{\boldsymbol{\chi}}F_{\theta}\,\overline{\delta\theta}+ \nabla_{\boldsymbol{\chi}}F_{\theta'}\,\overline{\delta\theta}'\right]\text{d}s+\left.\left[\widetilde{\delta s} \nabla_{\boldsymbol{\chi}}F\right]\right|_{s_1}^{s_2}\right]\cdot \delta\boldsymbol{\chi},\\
\delta^3 \mathcal{V}=&\int_{s_1}^{s_2}\widetilde{\delta s}'
\left[-3\, \widetilde{\delta\theta_s}^2\, F_{\theta'\theta'}-6 \,\widetilde{\delta\theta_s} \,\widetilde{\delta s}' F_{\theta'}+3\, \widetilde{\delta\theta}^2  F_{\theta\theta}+6\, \widetilde{\delta\theta} \,\nabla_{\boldsymbol{\chi}}F_\theta\cdot \delta\boldsymbol{\chi}\right]\text{d}s\\
&+\int_{s_1}^{s_2}\left[3\, \widetilde{\delta\theta}^2 \nabla_{\boldsymbol{\chi}}F_{\theta\theta}\cdot \delta\boldsymbol{\chi}+\widetilde{\delta\theta}^3 F_{\theta\theta\theta}\right]\text{d}s.
\end{split}
\end{equation}

 \paragraph{Proof.} 
 The general increment $\Delta \mathcal{V}$ for the functional $\mathcal{V}$ (\ref{generalfunctional}),
\begin{equation}
\label{genvariationApp}
\Delta \mathcal{V}=\int_{s_1^*}^{s_2^*}F\left[s^*,\theta^*(s^*),\frac{\partial\theta^*(s^*)}{\partial s^*},\boldsymbol{\chi}^*\right]\,\text{d}s^*-\int_{s_1}^{s_2}F\left[s,\theta(s),\theta'(s),\boldsymbol{\chi}\right]\,\text{d}s
\end{equation}
can be firstly transformed by evaluating the differential for the coordinate $s^*$ from Eq. (\ref{linexps})$_1$, namely
\begin{equation}
\text{d}s^*=(1+\epsilon \phi')\text{d}s.
\end{equation}
The following linear transformation of the Lagrange multipliers is introduced
\begin{equation}
\boldsymbol{\chi}^* = \boldsymbol{\chi}+\epsilon\textbf{x},
\end{equation}
where $\textbf{x}$ is independent of $s$, so that $\delta \boldsymbol{\chi}=\epsilon \textbf{x}$ 
denotes the vector of perturbations in the  Lagrange multipliers. 

The general variation $\Delta \mathcal{V}$ (\ref{genvariationApp}) can be  rewritten as
\begin{equation}
\label{diffV}
\Delta \mathcal{V}=\int_{s_1}^{s_2}\left\{F\left[s^*,\theta^*(s^*),\frac{\partial\theta^*(s^*)}{\partial s^*},\boldsymbol{\chi}^*\right](1+\epsilon \phi')-F\left[s,\theta(s),\theta'(s),\boldsymbol{\chi}\right]\right\}\,\text{d}s,
\end{equation}
so that its expansion  in the \lq small' parameter $\epsilon$ provides
\begin{equation}
\Delta \mathcal{V}\simeq \epsilon \,\mathcal{V}_1+\frac{\epsilon^2}{2}\mathcal{V}_2+\frac{\epsilon^3}{6} \mathcal{V}_3+o(\epsilon^3)
\end{equation}
where variations of the total potential energy are denoted by $\delta^i \mathcal{V}=\epsilon^i \mathcal{V}_i$. 

Since the vector $\boldsymbol{\chi}$ is assumed to be independent of $s$, the following relation holds
\begin{equation}
\frac{\text{d} F[s,\theta(s),\theta'(s),\boldsymbol{\chi}]}{\text{d} s}=F_s+F_\theta \theta'+F_{\theta'}\theta'',
\end{equation}
where the terms $F_\xi$ denote differentiations of the functional $F$ with respect to the generic scalar field $\xi$, and the following expression for the first variation $\delta \mathcal{V}$ is obtained
\begin{equation}
\label{1vartrue}
\delta \mathcal{V}=\int_{s_1}^{s_2} \left[\nabla_{\boldsymbol{\chi}}F\cdot \delta\boldsymbol{\chi}+F_\theta \widetilde{\delta \theta}+F_{\theta'}\widetilde{\delta \theta_s}+F\widetilde{\delta s}'+F_s\widetilde{\delta s} \right]\text{d}s,
\end{equation}
so that integration by parts leads to
\begin{equation}
\delta \mathcal{V}=\int_{s_1}^{s_2} \left[\nabla_{\boldsymbol{\chi}}F\cdot \delta\boldsymbol{\chi}+F_\theta\,\widetilde{\delta \theta}+F_{\theta'}\,\widetilde{\delta \theta_s}+F_s\widetilde{\delta s} \right]\text{d}s+\left.F\widetilde{\delta s}\right|_{s_1}^{s_2}-\int_{s_1}^{s_2} \widetilde{\delta s}\left[F_s+F_\theta \theta'+F_{\theta'}\theta''\right]\text{d}s,
\end{equation}
where $\nabla_{\boldsymbol{\chi}}F$ is the first gradient of $F$ taken with respect to the Lagrange multipliers and where the terms $F_s\widetilde{\delta s}$ within the integrals simplify. By substituting Eqs. (\ref{expansions})$_1$ and (\ref{truedervars})$_1$, many terms simplify and relation (\ref{1varcomplete}) is obtained (see also a similar proof by Gelfand and Fomin\cite{gelfand}),
which is the same  expression that  would be obtained by performing variations with respect to the perturbation $\overline{\delta\theta}$ introduced by Elsgolts \cite{elsgolts}, see Eq. (\ref{perturba2}).

The  expression for the general second variation $\delta^2 \mathcal{V}$ is obtained by  assuming  $\boldsymbol{\chi}$  independent of $s$ and  the functional $F$ linearly-dependent on $\boldsymbol{\chi}$ as
\begin{equation}
\label{2vtrue2}
\begin{split}
\delta^2 \mathcal{V}=&\int_{s_1}^{s_2}\left[\widetilde{\delta\theta_s}^2 F_{\theta'\theta'}+\widetilde{\delta\theta}^2 F_{\theta\theta}+\widetilde{\delta s}^2 F_{ss}+2\,\widetilde{\delta\theta}\,\widetilde{\delta\theta_s} F_{\theta\theta'}+2\,\widetilde{\delta s}\,\widetilde{\delta\theta_s} F_{s\theta'}+2\,\widetilde{\delta s}\,\widetilde{\delta\theta} F_{s\theta}\right]\text{d}s\\
&+ 2\int_{s_1}^{s_2}\widetilde{\delta s}' \left(\widetilde{\delta s}  F_s+\widetilde{\delta\theta}  F_\theta\right)\text{d}s+2\left[\int_{s_1}^{s_2}\left[\widetilde{\delta s} \nabla_{\boldsymbol{\chi}}F_{s}+\widetilde{\delta\theta} \nabla_{\boldsymbol{\chi}}F_{\theta}+ \widetilde{\delta\theta_s} \nabla_{\boldsymbol{\chi}}F_{\theta'}+\widetilde{\delta s}' \nabla_{\boldsymbol{\chi}}F\right]\text{d}s\right]\cdot\delta\boldsymbol{\chi}.
\end{split}
\end{equation}
where due to the linearity of $F$ with respect to the vector $\boldsymbol\chi$, the condition $\nabla_{\boldsymbol{\chi}\boldsymbol{\chi}}F=\textbf{0}$ holds true, being $\nabla_{\boldsymbol{\chi}\boldsymbol{\chi}}F$  the second gradient of $F$ taken with respect to the Lagrange multipliers.

Recalling Eq. (\ref{dirichconds}) and the isoperimetric constraints, and performing integration by parts, 
after several passages  the second variation $\delta^2 \mathcal{V}$ can be expressed  in terms of the variation $\overline{\delta\theta}$ as expression (\ref{23varcomplete})$_1$, 
which is the same  expression that would have been obtained by performing variations with respect to the perturbation $\overline{\delta\theta}$ from the beginning.

Finally, by considering the functional $F$ independent of $s$, the  third variation $\delta^3 \mathcal{V}$  follows as  Eq. (\ref{23varcomplete})$_2$.

\paragraph{General variational problem  for the \lq \emph{elastica sling}'.}
\label{specificvars}

With reference to the \lq \emph{elastica sling}', the functional $\mathcal{V}$ is given by the  total potential energy expressed by  Eq. (\ref{potentialsling}) and the isoperimetric constraints (\ref{constraints1}) imply that the perturbation in the rotation field $\widetilde{\delta\theta}$ satisfies Eqs. (\ref{dirichconds}), namely $\widetilde{\delta\theta}(s_1)=0,\,\widetilde{\delta\theta}(s_2)=0$.

Since a uniform bending stiffness $B$ is assumed, the functional $F$ in Eq. (\ref{generalfunctional}) is not explicitly dependent on the curvilinear coordinate $s$, therefore $F_s=0$. The expression for first variation of the total potential energy can be easily recovered from Eq. (\ref{1varcomplete}) as
\begin{equation}
\label{firstvar2}
\begin{array}{lll}
\delta\mathcal{V}=&\ds-\int_{s_{1}}^{s_{2}}\left[B\theta''(s)+R_x\sin\theta(s)-R_y\cos\theta(s)\right]\widetilde{\delta\theta}(s)\,\text{d}s\\[2mm]
&\ds-\left\{\frac{B \left[\theta'(s_2)\right]^2}{2}-R_x\cos\theta(s_2)-R_y\sin\theta(s_2)\right\}\left(\widetilde{\delta s}_2-\widetilde{\delta s}_1\right)\\[2mm]
&\ds-\delta R_x\left(d-\int_{s_1}^{s_2}\cos\theta(s)\,\text{d}s\right)+\delta  R_y\int_{s_1}^{s_2}\sin\theta(s)\,\text{d}s,
\end{array}
\end{equation}
where for the sake of brevity, the vanishing terms involving the multipliers $M_1$ and $M_2$ and related variations are not reported. By replacing the perturbation $\widetilde{\delta\theta}$ (for which the variational formulation is not affected by \lq pure-sliding' motions) with $\overline{\delta\theta}$, expression (\ref{firstvar2}) provides  Eq. (\ref{1varsling}).

Considering that the properties $F_{\theta\theta'}=F_s=\nabla_{\boldsymbol{\chi}}F_{\theta'}=\textbf{0}$ hold true for the \lq \emph{elastica sling}',the second variation can be obtained from the generic expression (\ref{23varcomplete})$_1$  as
\begin{equation}
\label{secondvar}
\begin{split}
\delta^2 \mathcal{V}=&\int_{s_1}^{s_2}\left[B\,\overline{\delta\theta}'^2-\overline{\delta\theta}^2 \left(R_x\cos\theta+R_y\sin\theta\right)\right]\,\text{d}s-\left.\left[B\theta'\theta''\right]\widetilde{\delta s}^2\right|_{s_1}^{s_2}\\
&+2\,\delta R_x\left(\int_{s_1}^{s_2}-\sin\theta\,\overline{\delta\theta}\,\text{d}s+\left.\cos\theta\widetilde{\delta s}\right|_{s_1}^{s_2}\right)+2\,\delta R_y\left(\int_{s_1}^{s_2}\cos\theta\,\overline{\delta\theta}\,\text{d}s+\left.\sin\theta\widetilde{\delta s}\right|_{s_1}^{s_2}\right),
\end{split}
\end{equation}
and the third variation from equation (\ref{23varcomplete})$_2$  as
\begin{equation}
\label{thirdvar}
\begin{split}
\delta^3 \mathcal{V}=&-\int_{s_1}^{s_2}B\,\theta''\,\overline{\delta\theta}^3\,\text{d}s-\left.\frac{B}{2}\widetilde{\delta s}^3\left[\frac{\theta''^2}{2}+\theta'^4\right]\right|_{s_1}^{s_2}+3\,\delta R_x\left(\int_{s_1}^{s_2}-\cos\theta\,\overline{\delta\theta}^2\,\text{d}s-\left.\sin\theta\,\overline{\delta\theta}\,\widetilde{\delta s}\right|_{s_1}^{s_2}\right)\\
&+3\,\delta R_y\left(\int_{s_1}^{s_2}-\sin\theta\,\overline{\delta\theta}^2\,\text{d}s+\left.\cos\theta\,\overline{\delta\theta}\,\widetilde{\delta s}\right|_{s_1}^{s_2}\right).
\end{split}
\end{equation}

\subsection{Equivalence of equilibrium and  stability for rods constrained by one or two sliding sleeves}

In this section, the equivalence between the equilibrium and the stability of an Euler's elastica constrained by one and two sliding sleeves is discussed. In particular, the proof of equivalence is achieved by firstly considering the effect of the so called \lq pure-sliding' as previously reported in Sect. \ref{puresliding}. This particular combination of perturbations is related to a neutral equilibrium (namely all superior variations are null) as the new \lq perturbed' configuration outside the constraints remains equal to the initial one. For this reason, some changes in the coordinates are introduced in order to transform the current problem into a new \lq virtual system' having a single variable endpoint only. By changing the governing equations through the aforementioned transformations, the first and second variations write  as
\begin{equation}
\begin{split}\label{12var_red}
\delta \mathcal{V}=&-\int_{0}^{\mathcal{L}}\left(B\Theta''+R_x\sin\Theta-R_y\cos\Theta\right)\overline{\delta\Theta}\,\text{d}S-\left(\frac{B}{2}\Theta'(\mathcal{L})^2-R_x\cos\Theta(\mathcal{L})-R_y\sin\Theta(\mathcal{L})\right)\widetilde{\delta \mathcal{L}},\\
\delta^2 \mathcal{V}=&\int_{0}^{\mathcal{L}}\left[B\,\overline{\delta\Theta}'^2-\overline{\delta\Theta}^2 \left(R_x\cos\Theta+R_y\sin\Theta\right)\right]\,\text{d}S-B\Theta'(\mathcal{L})\Theta''(\mathcal{L})\widetilde{\delta \mathcal{L}}^2\\
&+2\,\delta R_x\left(\int_{0}^{\mathcal{L}}-\sin\Theta\,\overline{\delta\Theta}\,\text{d}S+\cos\Theta(\mathcal{L})\widetilde{\delta \mathcal{L}}\right)+2\,\delta R_y\left(\int_{0}^{\mathcal{L}}\cos\Theta\,\overline{\delta\Theta}\,\text{d}S+\sin\Theta(\mathcal{L})\widetilde{\mathcal{L}}\right),
\end{split}
\end{equation}
where $\widetilde{\delta \mathcal{L}}=\widetilde{\delta s}_2-\widetilde{\delta s}_1$. The expressions for the first and second variations (\ref{12var_red}) respectively show  that the equilibrium and stability for the system with two sliding sleeves is coincident with those of the same system with one sliding sleeve replaced by a clamp, except for a neutral perturbation provided by a \lq pure-sliding'.

\paragraph{\lq Pure-sliding'.} A neutral equilibrium can be attained for the system with two sliding sleeves by introducing a constant perturbation in the curvilinear coordinate $\widetilde{\delta s}=\widetilde{\delta s}_1$,
such that 
the deformed configuration of the rod outside the constraints remains the same, namely $\theta^*(s^*)=\theta(s)$, and therefore because of Eq. (\ref{deltafunz}) the perturbation $\widetilde{\delta\theta}$ is null for any coordinate $s$
\begin{equation}
\label{simpshift}
\widetilde{\delta\theta}(s)=0\qquad\Rightarrow\qquad \overline{\delta\theta}(s)=-\theta'(s)\widetilde{\delta s}_1,
\end{equation}
but for non-null variations $\overline{\delta\theta}$ and $\widetilde{\delta s}_1$.

Moreover, due to the fact that both the equilibrium and perturbed solutions outside the constraints are the same, no difference can be obtained in terms of  derivatives, namely
\begin{equation}
\frac{\partial \theta^*(s^*)}{\partial s^*}=\frac{\partial \theta(s)}{\partial s},
\end{equation}
so that because of Eqs. (\ref{truedervars1}) and (\ref{truedervars}) the perturbation $\widetilde{\delta\theta_s}$ must vanish. Being the perturbation $\widetilde{\delta s}$ constant for any $s$, its derivative must obviously vanish ($\widetilde{\delta s}'=0$). Finally, no difference can be attained in terms of the reaction forces at both ends, so that $\delta\boldsymbol{\chi}=\textbf{0}$. Looking at equations (\ref{23varcomplete})$_1$ and (\ref{23varcomplete})$_2$ for the specific case of $F_s=0$, higher variations (at any order) are represented by complete forms (of the same order $i$ of the variation $\delta ^i \mathcal{V}$) of the aforementioned variations $\widetilde{\delta\theta}$, $\widetilde{\delta\theta_s}$, $\widetilde{\delta s}'$ and $\delta\boldsymbol{\chi}$, so that higher variations are all null ($\delta^n \mathcal{V}=0,\,\forall\, n\in \mathbb{N}$) in the case of \lq pure-sliding' and the equilibrium can be judged neutral along this specific direction in the perturbations' space. 

For these reasons, it is instrumental to exclude this \lq trivial' perturbation from the study of stability, thus focusing on the perturbations that cause a non null transformation of the equilibrium solutions. By means of an accurate mathematical proof, one can demonstrate that the stability of the considered elastica constrained by sliding sleeves at both ends can be studied with an equivalent elastica constrained with a single sliding sleeve, for instance located at the final end, while the other end is clamped.\footnote{A similar proof can be also performed by considering reverted constraints, so that one can conclude that the problem of a rod subject to sliding sleeves at both ends is equivalent (in terms of equilibrium configurations and stability) to a problem for which one of two ends is clamped.} 

Finally, it is worth to underline the fact that if the integrand of the  functional (\ref{generalfunctional}) would be explicitly dependent on $s$ (for instance, when the bending stiffness $B$ is  a given function of $s$ or  some loads are  applied to specific coordinates $s$ along the span of the rod), then all the variations of the total potential energy depend on the quantity $\widetilde{\delta s}$ (and not only on $\widetilde{\delta s}'$), therefore no invariance of the configuration and no vanishing of higher variations would be obtained under a  \lq pure-sliding'.

\paragraph{Proof of equivalence.} The proof of the aforementioned equivalence in terms of stability can be achieved by performing the following change of coordinates
\begin{equation}
\label{changecoord}
s-s_1=S,
\end{equation}
where $S\in [0,\mathcal{L}]$ is the new curvilinear coordinate of a \lq virtual' elastica having the initial end always coincident with the position of the first-sliding sleeve, so that it can be thought to represent a shorter rod with a clamped initial end and with a total \lq free' length $\mathcal{L}=s_2-s_1$ outside the two constraints. The rotation field of this new elastica with one variable endpoint only is denoted by $\Theta(S)$, so that the following relation holds
\begin{equation}
\label{thetaTheta}
\theta(s)=\Theta(S).
\end{equation}

Moreover, the generic transformed solution $\theta^*(s^*)$ is constrained to be the same holding for the clamped elastica, so that
\begin{equation}
\label{thetaThetastar}
\theta^*(s^*)=\Theta^*(S^*),
\end{equation}
where the following simple change of coordinates is introduced between the perturbed domains of the configurations outside the constraints
\begin{equation}
\label{sSstar}
S^*(s^*)=s^*-s_1^*,
\end{equation}
so that $S^*(s_1*)=0$.
These simple transformation rules allow for representing the perturbed solution of the rod constrained by two sliding sleeves in a new reference system for which the first end is fixed, but without changing the global deformed shape of both the equilibrium and perturbed configurations. Similarly to Eqs. (\ref{linexps}), (\ref{deltafunz}) and (\ref{pippo}), the following smooth transformations between the equilibrium and perturbed solutions of the virtually-clamped system are introduced
\begin{equation}
\label{transBigfunz}
S^*=S+\epsilon \Phi(S) +o(\epsilon),\qquad \Theta^*(S^*)=\Theta(S)+\epsilon \Psi(S) +o(\epsilon),
\end{equation}
together with the following transformation holding at a fixed coordinate $S$
\begin{equation}
\label{poppo}
\Theta^*(S)=\Theta(S)+\epsilon \overline\Psi(S) +o(\epsilon).
\end{equation}

The Eqs. (\ref{changecoord}) and (\ref{sSstar}) can be therefore substituted in the difference $\widetilde{\Delta s}$, see Eq. (\ref{deltafunz})$_1$, so that through Eq. (\ref{transBigfunz})$_1$ one obtains
\begin{equation}
\widetilde{\Delta s}=S^*+s^*_1-S-s_1=\epsilon \Phi(S)+\epsilon \phi(s_1) +o(\epsilon),\qquad\text{where}\qquad\widetilde{\Delta s} = \epsilon \phi(s)+o(\epsilon).
\end{equation}
Due to Eq. (\ref{truevars})$_1$, one obtains the following linear parts of the increments above
\begin{equation}
\label{deltas}
\widetilde{\delta s}=\widetilde{\delta S}+\widetilde{\delta s}_1,
\end{equation}
where the new variation $\widetilde{\delta S}=\epsilon\Phi$ is introduced and for which the following properties hold
\begin{equation}
\label{changecoord4}
\widetilde{\delta S}(s_1)=0,\quad\widetilde{\delta S}'=\widetilde{\delta s}',\quad\widetilde{\delta s}_2-\widetilde{\delta s}_1=\widetilde{\delta \mathcal{L}}.
\end{equation}
It is worth to underline that a \lq pure-sliding' of the rod 
like the one represented in Fig. \ref{fig_shiftfranz} (for which under $\widetilde{\delta s}=\text{const}=\Delta s_1$)
would cause the perturbation $\widetilde{\delta S}$ of the \lq virtually-clamped ' rod to be null, so that this new system is capable of catching the perturbations related to a change in the deformed configuration, but refined from those related to \lq pure-sliding'.

\begin{figure}[!h]
\renewcommand{\figurename}{\footnotesize{Fig.}}
    \begin{center}
   \includegraphics[width=0.8\textwidth]{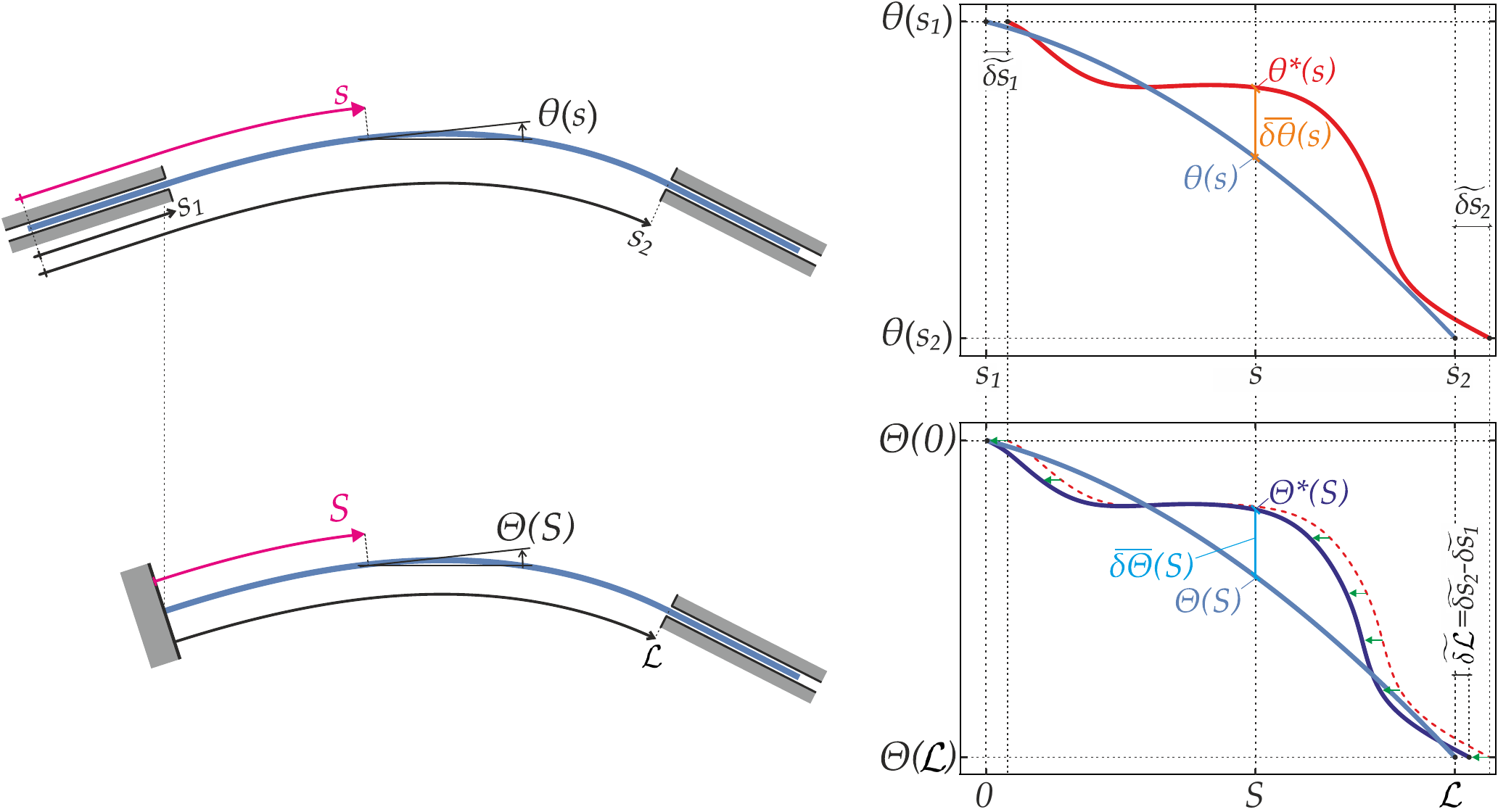}
    \caption{\footnotesize{(Left) The two equivalent systems of an elastica constrained with two (above) and one sliding sleeve (bottom). (Right, above) Equilibrium (blue) and perturbed (red) solutions for the rotation field $\theta(s)$ of the elastica constrained with two sliding sleeves and related perturbations $\widetilde{\delta s}_1$ and $\widetilde{\delta s}_2$ at the variable ends. (Right, bottom) Equilibrium (blue) and perturbed (dark blue) solutions for the rotation field $\Theta(S)$ of the clamped elastica constrained with one sliding sleeve at the final end; a unique perturbation $\widetilde{\delta \mathcal{L}}$ is considered at the final endpoint. Note the shifting of the perturbed field $\widetilde{\delta \theta}$ of a factor $\widetilde{\delta s}_1$.}
		}
    \label{fig_2sys}
    \end{center}
\end{figure}

An analogous calculation can be performed by substituting Eqs. (\ref{thetaTheta}) and (\ref{thetaThetastar}) in the difference $\widetilde{\Delta \theta}$, see Eq. (\ref{deltafunz})$_2$, thus obtaining
\begin{equation}
\label{cacca}
\widetilde{\Delta \theta}(s)=\Theta^*(S^*)-\Theta(S).
\end{equation}
By considering Eq. (\ref{transBigfunz})$_2$, the equation above can be therefore expanded for a \lq small' $\epsilon$, so that one obtains an analogous relation holding for the (virtually) clamped rod
\begin{equation}
\label{pertNew}
\widetilde{\delta \Theta}=\Theta'(S)\widetilde{\delta S}+\overline{\delta\Theta},
\end{equation}
where due to the identity between the differentials $dS=ds$, \lq$\,'\,$' denotes differentiation with respect to both coordinates $S$ or $s$. Moreover the following definitions have been introduced
\begin{equation}
\widetilde{\delta \Theta}=\epsilon \Psi,\qquad \overline{\delta \Theta}=\epsilon \overline\Psi,
\end{equation}
but due to identity between Eq. (\ref{deltafunz})$_2$ and (\ref{cacca}), the following conditions hold true
\begin{equation}
\widetilde{\delta\theta}(s)=\widetilde{\delta\Theta}(S),\qquad \overline{\delta \theta}\neq \overline{\delta \Theta},
\end{equation}
namely the \lq general' variations of the deformed configurations are identical for both systems, while the variations at fixed coordinates are connected by the following fundamental rule
\begin{equation}
\label{Hhrel}
\overline{\delta \Theta}=\overline{\delta \theta}+\theta'\widetilde{\delta s}_1.
\end{equation}
namely the variation at fixed coordinate $S$ of the new system is obtained by shifting the original one of a factor $\theta'\widetilde{\delta s}_1$, which is in turn of the same form of the one holding in the case of \lq pure-sliding' (see Eq. (\ref{simpshift})$_2$). The two considered systems are highlighted in Fig. \ref{fig_2sys} (left), together with (right) the rotation field $\theta(s)=\Theta(S)$ at equilibrium (highlighted in light blue) and the perturbed solution holding for the system with two sliding sleeves ($\theta^*(s)$, red) and one sliding sleeve ($\Theta^*(S)$, dark blue), where the latter is shifted horizontally from the right to the left of a factor $\delta s_1$ with respect to the former.\footnote{The case of a clamped final end can be studied by shifting $\Theta^*(S)$ of a factor $\widetilde{\delta s}_2$ in the opposite direction.}

In particular, due to vanishing of the perturbations at both ends\footnote{Due to the fact that $\widetilde{\delta S}(s_1)=0$, the variation at the initial end corresponds to the one holding for fixed coordinates, namely $\widetilde{\delta\Theta}(0)=\overline{\delta\Theta}(0)$.}
\begin{equation}
\widetilde{\delta\theta}(s_1)=\widetilde{\delta\Theta}(0)=0,\qquad \widetilde{\delta\theta}(s_2)=\widetilde{\delta\Theta}(L)=0,
\end{equation}
Eqs. (\ref{pertNew}) and (\ref{Hhrel}) lead to the following relations for the field $\overline{\delta \Theta}$
\begin{equation}
\overline{\delta \Theta}(0)=\overline{\delta \theta}(s_1)+\theta'(s_1)\widetilde{\delta s}_1,\qquad \overline{\delta \Theta}(L)=\overline{\delta \theta}(s_2)+\theta'(s_2)\widetilde{\delta s}_1,
\end{equation}
but due to Eqs. (\ref{expansions}), one finally obtains
\begin{equation}
\label{BCsclamp}
\overline{\delta \Theta}(0)=0,\qquad \overline{\delta \Theta}(L)=-\theta'(s_2)\widetilde{\delta s}_2+\theta'(s_2)\widetilde{\delta s}_1=-\Theta'(L)\widetilde{\delta \mathcal{L}},
\end{equation}
and therefore
\begin{equation}
\overline{\delta \Theta}(0)=\widetilde{\delta\Theta}(0)=0,\qquad \overline{\delta \Theta}(\mathcal{L})\neq \widetilde{\delta \Theta}(\mathcal{L}),
\end{equation}
where the obtained conditions are consistent with the assumption of a clamped end at coordinate $S=0$ and a sliding sleeve at $S=\mathcal{L}$.

Finally, the difference in the derivatives highlighted in Eq. (\ref{truedervars1}) can be rewritten in terms of the new perturbed solution, so that due to Eqs. (\ref{thetaTheta}) and (\ref{thetaThetastar}) and due to the identity $dS=ds$ between differentials of the two curvilinear coordinates one can write
\begin{equation}
\label{identityD}
\widetilde{\Delta \theta_s}= \frac{\partial \Theta^*(S^*)}{\partial s^*}-\frac{\partial \theta(s)}{\partial s}=\frac{\partial \Theta^*(S^*)}{\partial S^*}\frac{\partial S^*}{\partial s^*}-\frac{\partial \Theta(S)}{\partial S},
\end{equation}
where $\partial S^*/\partial s^*=1$. Following Gelfand and Fomin\cite{gelfand} (paragraph 37.3), one can write
\begin{equation}
\label{deltaUS}
\widetilde{\Delta \theta_s}=\frac{\partial\left[\Theta^*(S^*)-\Theta(S^*)\right]}{\partial S^*}+\frac{\partial\left[\Theta(S^*)-\Theta(S)\right]}{\partial S}+\left(\frac{\partial}{\partial S^*}-\frac{\partial}{\partial S}\right)\Theta(S^*),
\end{equation}
so that, after long iterations, one obtains the following formulation for the last term in the latter equation, namely
\begin{equation}
\label{EE3}
\left(\frac{\partial}{\partial S^*}-\frac{\partial}{\partial S}\right)\Theta(S^*)=-\epsilon\Phi'\Theta'.
\end{equation}
The first term in Eq. (\ref{deltaUS}) can be rewritten through Eq. (\ref{poppo}) as
\begin{equation}
\label{EE1}
\frac{\partial\left[\Theta^*(S^*)-\Theta(S^*)\right]}{\partial S^*}=\frac{\partial\epsilon\overline{\Psi}(S^*)}{\partial S^*}\simeq \epsilon \frac{\partial\left[\overline{\Psi}(S)+\epsilon\Phi\overline{\Psi}'(S)\right]}{\partial S}\frac{1}{1+\epsilon \Phi'},
\end{equation}
while the second term in Eq. (\ref{deltaUS}) can be expanded in Taylor series 
\begin{equation}
\label{EE2}
\frac{\partial\left[\Theta(S^*)-\Theta(S)\right]}{\partial S}\simeq \frac{\partial\left[\Theta'(S)(S^*-S)+o(S^*-S)\right]}{\partial S},
\end{equation}
so that merging the terms (\ref{EE3}), (\ref{EE1}) and (\ref{EE2}), after long iterations one obtains
\begin{equation}
\widetilde{\Delta\theta_s}=\epsilon \overline{\Psi}'+\epsilon\Theta''\Phi+o(\epsilon).
\end{equation}
By denoting with $\widetilde{\delta \Theta_S}$ the linear part of the previous difference, one obtains
\begin{equation}
\widetilde{\delta \Theta_S}=\overline{\delta\Theta}'+\Theta''\widetilde{\delta S},
\end{equation}
which is necessarily equal to $\widetilde{\delta \theta_s}$ due to identity between Eqs. (\ref{truedervars}) and (\ref{identityD}) and where due to Eq. (\ref{Hhrel}) the following property holds
\begin{equation}
\overline{\delta\Theta}'=\overline{\delta\theta}'+\Theta''\widetilde{\delta s}_1.
\end{equation}

The aforementioned formulations therefore provide the proof of the identity between \lq general' perturbations holding for the system constrained by two sliding sleeves, namely $\widetilde{\delta\theta}$, $\widetilde{\delta \theta_s}$ and the derivative $\widetilde{\delta s}'$, with those holding for a virtual system that has a single variable endpoint only, here denoted by $\widetilde{\delta\Theta}$, $\widetilde{\delta \Theta_S}$ and $\widetilde{\delta S}'$, respectively, which is in turn realized by a simple transformation of the reference system.\footnote{Note that the value assumed by vector $\delta\boldsymbol{\chi}=\epsilon\textbf{x}$ is independent on $s$, so that it can not be changed by transformations of the curvilinear coordinate.} By substituting the aforementioned perturbations within the functionals introduced in Sections \ref{generalvars} and \ref{specificvars}, one can express the stability of the system constrained by two sliding sleeves with an equivalent one written in terms of a set of perturbations that do not take into account variations at one of the two ends. Moreover, these perturbations are equivalent to those that one would obtain by considering, since the beginning, a system constrained with a single sliding sleeve while the second end is clamped. As previously reported, this condition holds true under the hypothesis of $F$ independent on $s$, otherwise no match can be in general obtained between the two systems.  

The formulation of the first and second variations for the virtually-clamped system can also be obtained by performing the substitution (\ref{Hhrel}) within the formulations holding for the rod with two sliding sleeves and expressed in terms of the variation $\overline{\delta\theta}$, Eqs. (\ref{firstvar2}) and (\ref{secondvar}). The first variation becomes
\begin{equation}
\begin{split}
\delta \mathcal{V}=&-\int_{0}^{\mathcal{L}}\left(B\Theta''+R_x\sin\Theta-R_y\cos\Theta\right)\overline{\delta\Theta}\,\text{d}S+\widetilde{\delta s}_1\int_{0}^{\mathcal{L}}\left(B\Theta''+R_x\sin\Theta-R_y\cos\Theta\right)\Theta'\,\text{d}S\\
&-\left.\left(\frac{B}{2}\Theta'^2-R_x\cos\Theta-R_y\sin\Theta\right)\right|_{\mathcal{L}}\widetilde{\delta s}_2+\left.\left(\frac{B}{2}\Theta'^2-R_x\cos\Theta-R_y\sin\Theta\right)\right|_{0}\widetilde{\delta s}_1,
\end{split}
\end{equation}
and, due to the  property $\Theta''\Theta'=(\Theta'^2/2)'$ and the integration by parts of the second integral,  the  expression  (\ref{12var_red})$_1$ of the transformed first variation is obtained.
The present formulation is therefore totally equivalent to the one holding in the case of one clamped end and performed through perturbations in the form (\ref{perturba2}) and (\ref{pippo}) as proposed by Elsgolts \cite{elsgolts}. 

Concerning the second variation, Eqs. (\ref{changecoord}), (\ref{thetaTheta}), (\ref{Hhrel}) and (\ref{changecoord4}) can be substituted in Eq. (\ref{secondvar}), thus obtaining 
\begin{equation}
\begin{split}
\delta^2 \mathcal{V}=&\int_{0}^{\mathcal{L}}\left[B\left(\overline{\delta\Theta}'-\Theta''\widetilde{\delta s}_1\right)^2-\left(\overline{\delta\Theta}-\Theta'\widetilde{\delta s}_1\right)^2 \left(R_x\cos\Theta+R_y\sin\Theta\right)\right]\,\text{d}S\\
&+2\,\delta R_x\left(\int_{0}^{\mathcal{L}}\left[-\sin\Theta\left(\overline{\delta\Theta}-\Theta'\widetilde{\delta s}_1\right)\right]\,\text{d}S+\cos\Theta(\mathcal{L})\widetilde{\delta s}_2-\cos\Theta(0)\widetilde{\delta s}_1\right)\\
&+2\,\delta R_y\left(\int_{0}^{\mathcal{L}}\left[\cos\Theta\left(\overline{\delta\Theta}-\Theta'\widetilde{\delta s}_1\right)\right]\,\text{d}S+\sin\Theta(\mathcal{L})\widetilde{\delta s}_2-\sin\Theta(0)\widetilde{\delta s}_1\right)\\
&-\left[B\Theta'(\mathcal{L})\Theta''(\mathcal{L})\right]\widetilde{\delta s}_2^2+\left[B\Theta'(0)\Theta''(0)\right]\widetilde{\delta s}_1^2.
\end{split}
\end{equation}

After several passages,  the  expression (\ref{12var_red})$_2$ for the second variation is obtained, 
which is coincident with  the second variation that  would be obtained by considering a system with just one variable endpoint since the beginning. This result provides a further proof of the equivalence between the stability property of elasticae constrained by one or two sliding sleeves.

\section{Mathematical details for the derivation of Eq. (\ref{omegapsi3})}\label{mathdet}
\setcounter{equation}{0}
\setcounter{figure}{0}

Considering the properties (\ref{tensprop}), the  last  term on the right-hand side term of Eq. (\ref{omegapsi1}) can be simplified as
 \begin{equation}
 \label{lastterm}
 H_1\left[ (\textbf{p}'\cdot \textbf{u})(\textbf{p}\cdot \textbf{u}') -(\textbf{p}'\cdot \textbf{u}')(\textbf{p}\cdot \textbf{u})\right]=H_1 \,\textbf{p}'\otimes\textbf{p}:\left(\textbf{u}\otimes\textbf{u}'-\textbf{u}'\otimes\textbf{u}\right),
 \end{equation}
 where the term $\left(\textbf{u}\otimes\textbf{u}'-\textbf{u}'\otimes\textbf{u}\right)$ on the right hand side is a skew-symmetric tensor. Due to the property (\ref{tensprop})$_2$, Eq. (\ref{lastterm}) can be written by  considering only the skew part of tensor $\textbf{p}'\otimes\textbf{p}$, thus
 \begin{equation}
H_1\left[ (\textbf{p}'\cdot \textbf{u})(\textbf{p}\cdot \textbf{u}') -(\textbf{p}'\cdot \textbf{u}')(\textbf{p}\cdot \textbf{u})\right]=H_1 \,\frac{\textbf{p}'\otimes\textbf{p}-\left(\textbf{p}'\otimes\textbf{p}\right)^\intercal}{2}:\left(\textbf{u}\otimes\textbf{u}'-\textbf{u}'\otimes\textbf{u}\right).
 \end{equation}
Moreover, because $\textbf{u}(s_1)=\textbf{0}$, the following relation holds
\begin{equation}\label{eqnapp0}
H_1 \left(\textbf{u}\otimes\textbf{u}'-\textbf{u}'\otimes\textbf{u}\right)=\int_{s_1}^{s}\left[H_1 \left(\textbf{u}\otimes\textbf{u}'-\textbf{u}'\otimes\textbf{u}\right)\right]'\text{d}\sigma,
\end{equation}
so that, by considering Eq. (\ref{psioper}) and the associativity property of the outer product
\begin{equation}
\left(\textbf{a}+\textbf{b}\right)\otimes\textbf{c}=\textbf{a}\otimes\textbf{c}+\textbf{b}\otimes\textbf{c},\quad \textbf{a}\otimes\left(\textbf{b}+\textbf{c}\right)=\textbf{a}\otimes\textbf{b}+\textbf{a}\otimes\textbf{c},
\end{equation}
summing and subtracting the term  $H_2 \textbf{u}\otimes\textbf{u}$ inside the integral operator of Eq. (\ref{eqnapp0}) finally provides
\begin{equation}
H_1 \left(\textbf{u}\otimes\textbf{u}'-\textbf{u}'\otimes\textbf{u}\right)=\int_{s_1}^{s}\left[\Psi(\textbf{u})\otimes \textbf{u}- \textbf{u}\otimes\Psi(\textbf{u})\right]\text{d}\sigma.
\end{equation}
Still considering $\textbf{u}(s_1)=\textbf{0}$, the first term on the right-hand side  of Eq. (\ref{omegapsi1}) can be rewritten as
\begin{equation}
\delta\theta \,\textbf{p}\cdot \Psi(\textbf{u})=
\left[\textbf{p}\otimes\textbf{p}:\int_{s_1}^{s}\Psi(\textbf{u})\otimes \textbf{u}\,\text{d}\sigma\,\right]'-\left(\textbf{p}\otimes\textbf{p}\right)':\int_{s_1}^{s}\Psi(\textbf{u})\otimes \textbf{u}\,\text{d}\sigma,
\end{equation}
so that, due to the property
\begin{equation}
\textbf{a}\otimes\textbf{b}=\left(\textbf{b}\otimes\textbf{a}\right)^\intercal,
\end{equation}
Eq. (\ref{omegapsi1})  can be rewritten as
\begin{equation}
\label{omegapsi2}
\begin{split}
\delta\,\theta\Psi(\delta\theta)=& H_1(\textbf{p}'\cdot \textbf{u})^2+\left[\textbf{p}\otimes\textbf{p}:\int_{s_1}^{s}\Psi(\textbf{u})\otimes \textbf{u}\,\text{d}\sigma-\omega\,H_1 (\textbf{p}'\cdot \textbf{u})\right]'\\
&+\,\frac{\textbf{p}'\otimes\textbf{p}-\left(\textbf{p}'\otimes\textbf{p}\right)^\intercal}{2}:\left[\int_{s_1}^{s}\left[\Psi(\textbf{u})\otimes \textbf{u}- \left(\Psi(\textbf{u})\otimes\textbf{u}\right)^\intercal\right]\text{d}\sigma\right]
\\
&-2 \,\frac{\textbf{p}'\otimes\textbf{p}+\left(\textbf{p}'\otimes\textbf{p}\right)^\intercal}{2}:\int_{s_1}^{s}\Psi(\textbf{u})\otimes \textbf{u}\,\text{d}\sigma.
\end{split}
\end{equation}
Finally, due to the property (\ref{tensprop})$_2$, the last term of the previous equation can be rewritten by considering only the symmetric part of the tensor inside the integral, thus 
\begin{equation}
2\,\frac{\textbf{p}'\otimes\textbf{p}+\left(\textbf{p}'\otimes\textbf{p}\right)}{2}:\int_{s_1}^{s}\Psi(\textbf{u})\otimes \textbf{u}\,\text{d}\sigma=\frac{\textbf{p}'\otimes\textbf{p}+\left(\textbf{p}'\otimes\textbf{p}\right)}{2}:\int_{s_1}^{s}\left[\Psi(\textbf{u})\otimes \textbf{u}+ \left(\Psi(\textbf{u})\otimes\textbf{u}\right)^\intercal\right]\text{d}\sigma,
\end{equation}
which, exploited  in Eq. (\ref{omegapsi2}), allows to achieve Eq. (\ref{omegapsi3}).

\setcounter{equation}{0}
\setcounter{figure}{0} 

\begin{thebibliography}{99}

\setlength{\itemsep}{-1.0mm}

\bibitem{abbassi}
Abbasi, A., Sano, T.G., Yan, D., Reis, P.M. (2023) Snap buckling of bistable beams under combined mechanical and magnetic loading. \emph{Philosophical Transactions of the Royal Society A: Mathematical, Physical and Engineering Sciences} 381 (2244), 20220029

\bibitem{alfalahi}
Alfalahi, H., Renda, F.,  Stefanini, C. (2020) Concentric tube robots for minimally invasive surgery: Current applications
and future opportunities. \emph{IEEE Transactions on Medical Robotics and Bionics} 2 (3), 410--424


\bibitem{amor}
Amor, A., Fernandes, A., Pouget, J., Maurini, C. (2023)
Nonlinear dynamics and snap-through regimes of a bistable buckled beam excited by an electromagnetic Laplace force.
\emph{European Journal of Mechanics - A/Solids} (98),
104834

\bibitem{armaniniconf}
Armanini, C., Dal Corso, F., Misseroni, D., Bigoni, D. (2019). Configurational forces and nonlinear structural dynamics.
\emph{Journal of the Mechanics and Physics of Solids}, 130, 82--100.

\bibitem{arakawa}
Arakawa, K., Giorgio-Serchi, F., Mochiyama, H. (2021) Snap Pump: A Snap-Through Mechanism for a Pulsatile Pump. IEEE Robotics and Automation Letters  6 (2),  803--810

\bibitem{ballarini}
Ballarini, R., Royer-Carfagni, G. (2016)
A Newtonian interpretation of configurational forces on dislocations and cracks. 
\emph{Journal of the Mechanics and Physics of Solids} 95, 602--620

\bibitem{bertoldirev}
Bertoldi, K., Vitelli, V., Christensen, J., van Hecke, M. (2017) Flexible mechanical metamaterials. \emph{Nature Reviews} 2: 17066.


\bibitem{bigoniblade}
Bigoni, D., Bosi, F., Dal Corso, F. and Misseroni, D. (2014)
Instability of a penetrating blade.
\emph{Journal of the Mechanics and Physics of Solids}, 64, 411--425.


\bibitem{bigoniconf}
Bigoni, D., Dal Corso, F., Bosi, F. and Misseroni, D. (2015)
Eshelby-like forces acting on elastic structures: theoretical and experimental proof. \emph{Mechanics of Materials} 80, 368--374.

\bibitem{bigonitorsional}
Bigoni, D., Dal Corso, F., Misseroni, D. and Bosi, F. (2014).
Torsional locomotion. \emph{Proceedings of the Royal Society A} 470, 20140599.


\bibitem{bordiga}
Bordiga, G., Cabras, L., Piccolroaz, A.,  Bigoni, D. (2019) Prestress Tuning of Negative Refraction and Wave Channeling from Flexural Sources. \emph{Applied Physics Letters} 114, 041901


\bibitem{bolza} Bolza, O. (1902). Proof of the sufficiency of Jacobi’s condition for a permanent sign of the second variation in the so-called isoperimetric problems. \emph{Transactions of the American Mathematical Society}, 3, 305--305.

\bibitem{born} Born, M. (1906). Untersuchungen \"uber die Stabilit\"at der elastischen Linie in Ebene und Raum, unter verschiedenen Grenzbedingungen. Dieterichsche Universit\"ats-Buchdruckerei, G\"ottingen.


\bibitem{bosiarmscale}
Bosi, F., Misseroni, D., Dal Corso, F. and Bigoni, D. (2014). 
An elastica arm scale.
\emph{Proceedings of the Royal Society A}, 470, 20140232.

\bibitem{bosiinj}
Bosi, F., Misseroni, D., Dal Corso, F. and Bigoni, D. (2015).
Development of configurational forces during the injection of an elastic rod
Extreme Mechanics Letters, 471 83-88.


\bibitem{bosirestab}
Bosi, F., Misseroni, D., Dal Corso, F., Neukirch, S., Bigoni, D. (2016).
Asymptotic self-restabilization of a continuous elastic structure.
Physical Review E, 94 (6), 063005.

\bibitem{calisti}
Calisti, M.  and Picardi, G.  and Laschi, C. (2017) Fundamentals of soft robot locomotion. \emph{Journal of The Royal Society Interface}, 14 (130), 20170101.

\bibitem{carta}
Carta, G., Nieves, M.J., Brun, M. (2023)
Forcing the silence of the Lamb waves: Uni-directional propagation in structured gyro-elastic strips and networks. \emph{
European Journal of Mechanics - A/Solids}  101, 105070 

\bibitem{cazzolli}
Cazzolli, A., Dal Corso, F.  (2018)
Snapping of elastic strips with controlled ends.
\emph{Int. J. Sol. Struct.}, 162, 285--303.

\bibitem{cazzolliECM}
Cazzolli, A., Misseroni, D., Dal Corso, F.  (2018)
Elastica catastrophe machine: theory, design and experiments.
\emph{J. Mech. Phys. Sol.}, 136: 103735.

\bibitem{dalcorsochannel}
Dal Corso, F., Misseroni, D., Pugno, N.M., Movchan, A.B., Movchan, N.V., Bigoni, D. (2017) Serpentine locomotion through elastic energy release.
\emph{Journal of the Royal Society Interface} 14: 20170055.

\bibitem{dalcorsonested}
Dal Corso, F., Tallarico, D., Movchan, N., Movchan, A., Bigoni, D. (2019)
Nested Bloch waves in elastic structures with configurational forces.
\emph{Philosophical Transactions of the Royal Society A}, 377: 20190101.

\bibitem{elsgolts}
Elsgolts, L. (1977).
\emph{Differential equations and the calculus of variations}. Mir Publishers

\bibitem{eshelby}
Eshelby, J.D. Energy relations and the energy-momentum tensor in continuum mechanics. In \emph{Fundamental contributions to the continuum theory of evolving phase interfaces in solids}. Springer, 1999, 82--119.

\bibitem{filipov} Filipov, E.T., Tachi, T., Paulino, G.H. (2015) Origami tubes assembled into stiff, yet reconfigurable structures and metamaterials. \emph{Proceedings of the National Academy of Sciences} 112 (40), 12321.

\bibitem{frish}
Frisch-Fay, R. (1962).
\emph{Flexible bars}. Butterworths

\bibitem{garau}
Garau, M., Carta, G., Nieves, M.J., Jones,  I.S., Movchan, N.V., Movchan, A.B. (2018) Interfacial waveforms in chiral lattices with gyroscopic spinners. \emph{Proceedings of the Royal Society A} 474: 20180132

\bibitem{gelfand}
Gelfand, I.M., Fomin, S.V. (1963).
\emph{Calculus of Variations}. Prentice-Hall


\bibitem{goldberg}
Goldberg, N.N., O’Reilly, O.M. (2022) A material momentum balance law for shells and plates with application to phase transformations and adhesion. \emph{Acta Mech} 233, 3535--3555

\bibitem{Han2022}
Han, S. (2022) 
Configurational forces and geometrically exact formulation of sliding beams in non-material domains. \emph{Computer Methods in Applied Mechanics and Engineering} 395: 115063

\bibitem{Han2023a}
Han, S. (2023) Configurational forces and ALE formulation for geometrically exact, sliding shells in non-material domains. \emph{Computer Methods in Applied Mechanics and Engineering} 412: 116106

\bibitem{Han2023b}
Han, S., Bauchau, O.A. (2023) 
Configurational forces in variable-length beams for flexible multibody dynamics. \emph{Multibody System Dynamics} 58(3-4), 275--298

\bibitem{kochmann}
Kochmann, D., Bertoldi, K. (2017) Exploiting Microstructural Instabilities in Solids and
Structures: From Metamaterials to Structural Transitions. \emph{Applied Mechanics Review} 69, 050801

\bibitem{koutso}
Koutsogiannakis, P., Misseroni, D., Bigoni, D., Dal Corso, F. (2023)
Stabilization against gravity and self-tuning of an elastic variable-length rod through an oscillating sliding sleeve.
\emph{Journal of the Mechanics and Physics of Solids}, 181: 105452.

\bibitem{liakou}
Liakou, A., Detournay, E. (2018) Constrained buckling of variable length elastica: Solution by geometrical segmentation. \emph{International Journal of Non-Linear Mechanics} 99, 204--217

\bibitem{liu} 
Liu, K., Pratapa, P.P., Misseroni, D., Tachi, T.,  Paulino, G.H. (2022). Triclinic metamaterials by tristable origami with reprogrammable frustration. \emph{Advanced Materials,
} 34 (43), 2107998. 

\bibitem{love} 
Love, A.E.H. (1927) \emph{A treatise on the mathematical theory of elasticity}. Cambridge University
Press.

\bibitem{mahoney} 
Mahoney, A.W.,  Gilbert, H.B., Webster III, R.J. A
review of concentric tube robots: modeling, control, design, planning,
and sensing, in \emph{The Encyclopedia of Medical Robotics}  Chapter 7, 181--202 (2018).

\bibitem{majidi}
Majidi, C., O'Reilly, O.M., Williams, J.A. (2012). On the stability of a rod adhering to a rigid surface: Shear-induced stable adhesion and the instability of peeling. \emph{Journal of the Mechanics and Physics of Solids}, 60, 827--843.

\bibitem{maurini} 
Maurini, C., Pouget, J. Vidoli, S. (2007)
Distributed piezoelectric actuation of a bistable buckled beam.
\emph{European Journal of Mechanics - A/Solids} 26 (5),
837--853

\bibitem{misu0}
Misu, K., Yoshii, A., Mochiyama, H. (2018) A Compact Wheeled Robot that Can Jump while Rolling. \emph{2018 IEEE/RSJ International Conference on Intelligent Robots and Systems (IROS)}, Madrid, Spain, 7507--7512


\bibitem{misu}
Misu,  K., Mochiyama, H. (2021) Arched snap motor: power flow analysis, \emph{Advanced Robotics}, 35:18, 1107--1115

\bibitem{nadkarni}
Nadkarni, N., Arrieta, A.F., Chong, C., Kochmann, D.M., Daraio, C. (2016) Unidirectional transition waves in bistable lattices. \emph{Physical Review Letters} 116, 244501.

 
\bibitem{payne}
Payne-Gallway R. (1907) \emph{A summary of the history, construction and effects in warfare of the projectile-throwing engines of the ancients}. Longmans, Green and Co, London.

\bibitem{raney}
Raney, J.R., Nadkarni, N., Daraio, C., Kochmann, D.M., Lewis, J.A., Bertoldi, K. (2016) Stable
propagation of mechanical signals in soft media using stored elastic energy. 
\emph{Proceedings of the National Academy of Sciences}
113 (35), 9722--9727.

\bibitem{reis}
Reis P.M. (2015) A Perspective on the Revival of Structural (In) Stability With Novel Opportunities for Function: From Buckliphobia to Buckliphilia. \emph{ASME Journal Applied Mechanics} 82, 111001-1.

\bibitem{renda}
F Renda, C Messer, C Rucker, and F Boyer (2021) A sliding-rod variable-strain model for concentric tube robots. \emph{IEEE
Robotics and Automation Letters} 6 (2), 3451--3458

\bibitem{sano}
Sano, T.G., Wada, H. (2019) Twist-Induced Snapping in a Bent Elastic Rod and Ribbon. \emph{Physical Review Letters} 122 (11), 114301

\bibitem{venkatadri}
Venkatadri, T.K., Henzel, T., Cohen, T. (2023)
Torsion-induced stick-slip phenomena in the
delamination of soft adhesives
\emph{Soft Matter} 19: 2319

\bibitem{wang}
Wang, Z-Q.,  Detournay, E. (2022)
Eshelbian force on a steadily moving liquid blister. \emph{
International Journal of Engineering Science}  170: 103591

\bibitem{wen}
Wen, R., Wang, Z., Yi, J., Hu, Y. (2023) Bending-activated biotensegrity structure enables female Megarhyssa to cross the barrier of Euler’s critical force. \emph{Science Advances} 9 (42): eadi8284


\bibitem{yagi}
Yagi, K., Suzuki, K., Mochiyama, H. (2018) Human joint impedance estimation with a new wearable device utilizing snap-through buckling of closed-elastica. \emph{IEEE Robot Autom Lett.} 3 (3), 1506--1513

\bibitem{yamada} 
Yamada, A., Sugimoto, Y., Mameda, H., Fujimoto, H. (2012) An impulsive force generator based on
closed elastica with bending and twisting and its application to quick turning motion of swimming robot. \emph{Journal Robotics Soc. Japan} 29 (10), 923--933


\end{thebibliography}
\end{document}